\pdfoutput=1
\documentclass[a4paper,12pt]{article}
\usepackage{geometry}
\newgeometry{textwidth=0.8\paperwidth,top=0.07\paperheight,headheight=0.02\paperheight,headsep=0.03\paperheight,footskip=0.07\paperheight}

\usepackage[utf8]{inputenc}
\usepackage{CJKutf8}
\usepackage[font=small,labelfont=bf]{caption}
\usepackage{amsmath}
\usepackage{amsfonts}
\usepackage{multirow}
\usepackage{subcaption}
\usepackage{textcomp}
\usepackage{adjustbox}
\usepackage{booktabs}
\graphicspath{{fig/}}

\usepackage{enumitem}
\setlist[enumerate]{nosep}
\setlist[itemize]{nosep}
\setlist{labelindent=0em, leftmargin=\parindent}

\usepackage[affil-it]{authblk}
\usepackage{nameref}
\usepackage{float}
\usepackage{mathrsfs}
\usepackage{physics}
\usepackage[version=4]{mhchem}
\usepackage{siunitx}
\usepackage{multirow}
\def\secref#1{\ref{#1}~\nameref{#1}}

\begin{document}
\title{Reconstruction of Point Events in Liquid-Scintillator Detectors Subjected to Total Reflection}
\author[a,b,c]{Wei~Dou(\begin{CJK}{UTF8}{gbsn}窦威\end{CJK})}
\author[a,b,c,d,1]{Benda~Xu(\begin{CJK}{UTF8}{gbsn}续本达\end{CJK})\footnote{Corresponding author. orv@tsinghua.edu.cn}}
\author[e]{Jianfeng~Zhou(\begin{CJK}{UTF8}{gbsn}周建锋\end{CJK})}
\author[a,b,c]{Zhe~Wang(\begin{CJK}{UTF8}{gbsn}王\kern0.5em\hbox{\scalebox{0.5}[1]{吉}\scalebox{0.5}[1]{吉}}\end{CJK}\kern1.5em)}
\author[a,b,c]{Shaomin~Chen(\begin{CJK}{UTF8}{gbsn}陈少敏\end{CJK})}

\affil[a]{Department of Engineering Physics, Tsinghua University, Beijing, China}
\affil[b]{Center for High Energy Physics, Tsinghua University, Beijing, China}
\affil[c]{Key Laboratory of Particle \& Radiation Imaging (Tsinghua University), Ministry of Education, Beijing, China}
\affil[d]{Kavli Institute for the Physics and Mathematics of the Universe, UTIAS, the University of Tokyo, Tokyo, Japan}
\affil[e]{Xingfan Information Technology Co., Ningbo, China}
\maketitle

\begin{abstract}
    The outer water buffer is an economic option to shield the external radiative backgrounds for liquid-scintillator neutrino detectors. However, the consequential total reflection of scintillation light at the media boundary introduces extra complexity to the detector optics. This paper develops a precise detector-response model by investigating how total reflection complicates photon propagation and degrades reconstruction. We first parameterize the detector response by regression, providing an unbiased energy and vertex reconstruction in the total reflection region while keeping the number of parameters under control.  From the experience of event degeneracy at the Jinping prototype, we then identify the root cause as the multimodality in the reconstruction likelihood function, determined by the refractive index of the buffer, detector scale and PMT coverage. To avoid multimodality, we propose a straightforward criterion based on the expected photo-electron-count ratios between neighboring PMTs.  The criterion will be used to ensure success in future liquid-scintillator detectors by guaranteeing the effectiveness of event reconstruction.
\end{abstract}

\section{Introduction}
\label{sec: introduction}
Liquid-scintillator (LS) detectors, such as Borexino~\cite{2002science}, KamLAND~\cite{PhysRevLett.100.221803}, SNO+~\cite{Andringa_2016} and JUNO~\cite{JUNO:2021vlw}, obtain lower energy thresholds and better energy resolutions than water Cherenkov detectors by enhanced light yields. It contains rich contemporary physics topics, including neutrino mass ordering~\cite{an2016neutrino,noauthor_juno_2022}, neutrinoless double beta decay~\cite{PhysRevLett.117.082503, KamLAND-Zen:2022tow, SNO:2021xpa} as well as terrestrial, solar and supernova neutrinos~\cite{beacom_letter_2017}.

\begin{figure}
	\centering
	\includegraphics[width=0.5\linewidth, trim=120 20 250 20, clip,]{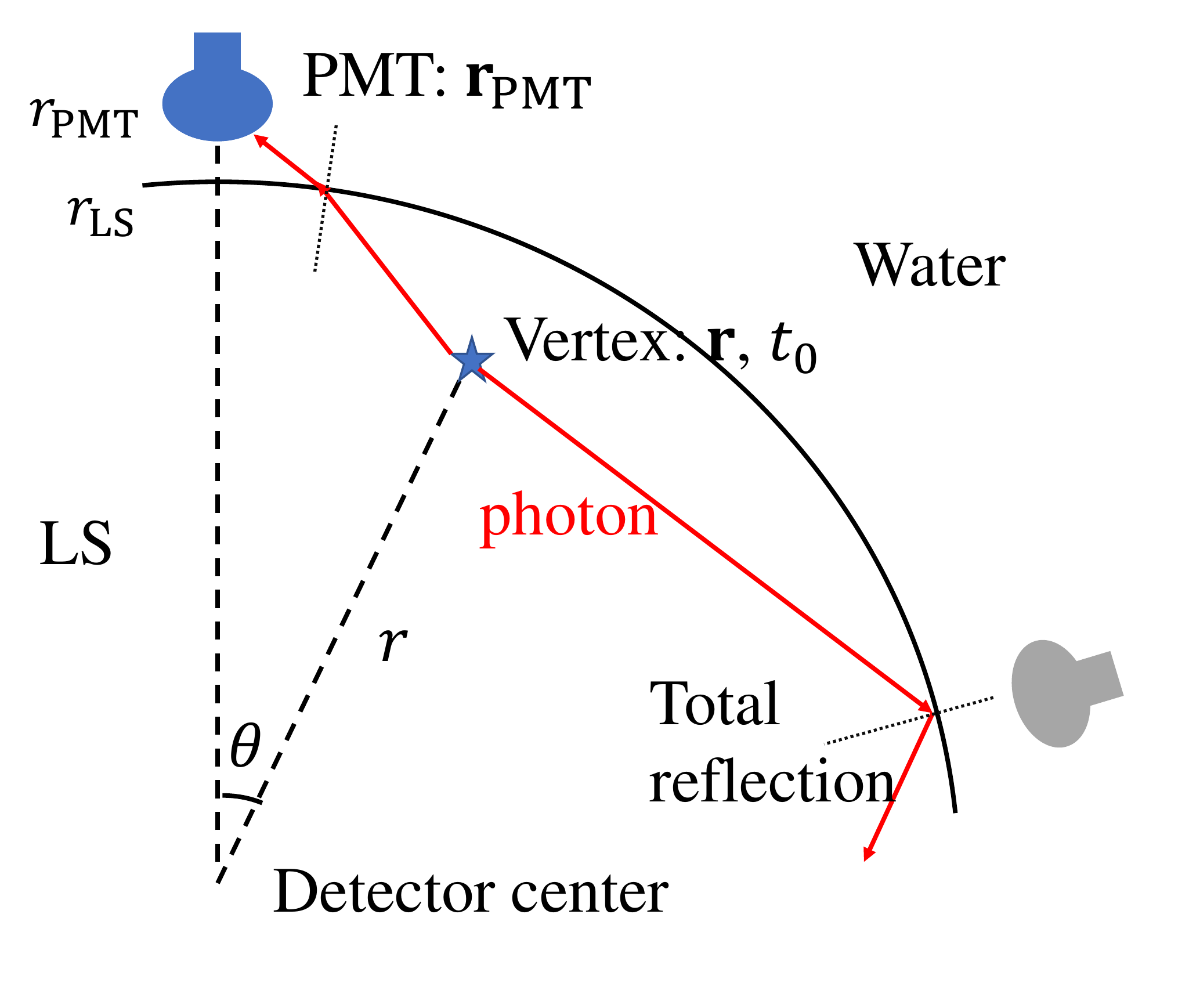}
	\caption{The particle deposits energy at position $\mathbf{r}$ on time $t_0$. The red lines are two typical photon tracks. Photons can reach the blue PMT directly but not the grey one due to TR. $r_\mathrm{LS}$ is the radius of the LS. The $i$-th PMT is located at position $\mathbf{r}_{\mathrm{PMT}, i}$ with $|\mathbf{r}_{\mathrm{PMT}, i}|= r_\mathrm{PMT}$. A function of $(\mathbf{r}, \mathbf{r}_{\mathrm{PMT}, i})$ can be simplified to that of $(r, \theta)$ by spherical symmetry, where $r = |\mathbf{r}|$ and $\theta$ is the central angle defined by the vertex and the PMT.}
	\label{fig: SH}
\end{figure}

The crucial step of energy reconstruction in LS detectors is to predict the \emph{photon electron} (PE) count and timing on each \emph{photomultiplier tube} (PMT) for a given vertex.  Vertex reconstruction thus influences energy resolution.  As illustrated in figure~\ref{fig: SH}, if an LS detector is surrounded by a water buffer, each PMT can be blind to photons in a specific region due to \emph{total reflection} (TR), causing a great challenge to vertex reconstruction.

The prevalent event reconstruction methods fall into 3 groups. \emph{Barycenter} (BC) averages all PMT positions weighted by PE. It is biased but fast~\cite{Liu_2018, Kim:2012cuq, li2021event}, often employed as the initial values to more advanced algorithms. \emph{Maximum likelihood estimation} (MLE) uses predicted PE~\cite{CHOOZ:2002qts} or timing~\cite{Liu_2018,Back_2012,Tajima:2003zz} or both~\cite{GALBIATI2006700,renocollaboration2010reno}. Its theoretical uncertainty is discussed by C. Galbiati et al~\cite{GALBIATI2006700}. The vertex reconstruction results are biased at the TR region if only use timing information~\cite{li2021event}. Li~\cite{li2021event} and Huang~\cite{Huang:2021baf} predict PE using interpolation with a map to achieve unbiased results but introduce a high degree of freedom. Machine learning methods~\cite{qian2021vertex} such as neural networks and decision trees have good modeling power to describe TR optics but need high-quality labeled training datasets only available from detector simulations.

In this paper, we develop an improved MLE-based event reconstruction resilient to TR. Section~\ref{sec2} formulates a detector-response model and tests it with Monte Carlo (MC) simulation data. Section~\ref{sec3} analyzes the event degeneracy caused by the multimodality in the likelihood function. Section~\ref{sec4} shows the reconstruction performance. Finally, section~\ref{sec5} discusses the extensibility and limitation of our method.  The symbol conventions for this paper are in table~\ref{tab:symbol}.

\begin{table}[!ht]
	\centering
	\caption{Definitions of symbols}
	\begin{tabular}{cll}
		\hline\hline
		variable                                & meaning (r.v. for random variable)            & first appearance in section                                      \\
		\hline
		$E, t_0$                                & visible energy and event start time           & \multirow{4}{*}{\secref{sec2}}                                   \\
		$\mathbf{r}, \mathbf{r}_{\mathrm{PMT}}$ & vertex and PMT's position                     &                                                                  \\
		$i, j$                                  & PMT index and hit index                       &                                                                  \\
		$r_\mathrm{LS}, r_\mathrm{PMT}$         & radii of the LS and PMT's position            &                                                                  \\
		\cline{3-3}
		$r=|\mathbf{r}|$                        & vertex radius                                 & \multirow{2}{*}{\secref{subsec: PE and timing prediction}}       \\
		$\theta$                                & central angle defined by PMT and vertex       &                                                                  \\
		\cline{3-3}
		$\hat{E}, \hat{\mathbf{r}}$             & reconstructed energy and position             & \secref{subsec: MC data results}                                 \\
		$\theta_\mathrm{acr}$                   & central angle defined by vertex and           & \secref{subsec: Other reflections by acrylic shell}              \\
		                                        & point of incidence to the acrylic shell       &                                                                  \\
		\hline
		$n_i$                                   & observed PE on $i$th PMT                      & \multirow{2}{*}{\secref{sec2}}                                   \\
		$t_{ij}$                                & observed $j$th hit's timing on $i$th PMT      &                                                                  \\
		\cline{3-3}
		$q_{ij}$                                & reconstructed $j$th hit's charge on $i$th PMT & \multirow{1}{*}{\secref{subsec: Raw data results}}               \\
		\cline{3-3}
		$N_\mathrm{PMT}$                        & number of PMTs                                & \secref{subsec: cosine distance}                                 \\
		$N_\mathrm{hit}$                        & total hit number                              & \secref{subsec: MC data results}                                 \\
		\hline
		$\lambda_i, \lambda_i(r, \theta)$       & predicted PE on $i$th PMT                     & \multirow{2}{*}{\secref{sec2}}                                   \\
		$T_i, T_i(r, \theta)$                   & predicted timing on $i$th PMT                 &                                                                  \\
		\cline{3-3}
		$c^\lambda_l, c^\lambda_l(r)$           & $l$th PE coefficient                          & \multirow{5}{*}{\secref{subsec: PE and timing prediction}}       \\
		$c^T_l, c^T_l(r)$                       & $l$th timing coefficient                      &                                                                  \\
		$c^\lambda_{l,m}, c^T_{l,m}$            & PE and timing coefficient                     &                                                                  \\
		$\mu_i, U_i$                            & PE and timing PMT-specific offset             &                                                                  \\
		$P_l$                                   & Legendre polynomial                           &                                                                  \\
		\hline
		$\psi(t)$                               & scintillation time profile                    & \secref{sec2}                                                    \\
		\cline{3-3}
		$R^0(t)$                                & timing PDF by $\psi(t)$ and TTS, etc.         &                                                                  \\
		$\mathscr{L}$                           & likelihood function                           & \multirow{4}{*}{\secref{subsec: PE and timing prediction}}       \\
		$\mathscr{R}_\tau, \mathscr{R}_\tau(t)$ & loss function using $\tau$-quantile           &                                                                  \\
		$R(t)$                                  & quantile regression approximation to $R^0(t)$ &                                                                  \\
		$\tau, t_\mathrm{s}$                    & quantile value and time scale                 &                                                                  \\
		\cline{3-3}
		$\tau_r, \tau_d$                        & rise and decay time constant                  & \secref{subsec: The training and validation sets}                \\
		$D_\mathrm{KL}$                         & Kullback-Leibler divergence                   & \secref{subsec: Raw data results}                                \\
		\hline
		$\Lambda$                               & PE pattern                                    & \multirow{2}{*}{\secref{subsec: cosine distance}}                \\
		$d_{\cos}$                              & cosine distance                               &                                                                  \\
		\cline{3-3}
		$C_{1r}(\mathbf{r})$                    & contour function                              & \multirow{3}{*}{\secref{subsec: Cosine distance for 3-PMT case}} \\
		$\Omega$                                & solid angle                                   &                                                                  \\
		$\beta$                                 & angle of incidence on a PMT                   &                                                                  \\
		\hline\hline
	\end{tabular}
	\label{tab:symbol}
\end{table}

\section{Detector response by regression}
\label{sec2}
This section develops a model to predict the PE and timing in an LS detector that is suitable for TR. The radius of the LS is $r_\mathrm{LS}$. Figure~\ref{fig: SH} shows how a typical LS detector works. An ionizing particle begins to deposit energy \(E\) at the position \(\mathbf{r}\) on time \(t_0\). It produces scintillation photons, each obeying a scintillation \emph{time profile} \(\mathrm{\psi}(t)\). PMTs are located at position $\mathbf{r}_{\mathrm{PMT}}$ with $|\mathbf{r}_{\mathrm{PMT}}|= r_\mathrm{PMT}$. A photon travels to a PMT with \emph{time of flight} (TOF) and induces a PE with the probability of \emph{quantum efficiency} (QE). The PE gets amplified by a series of dynodes inside the PMT with \emph{transit time} (TT), whose \emph{spread} is TTS. The whole process is called an \emph{event}.

The event is characterized by PE counts $n_i$ and timings $t_{ij}$ extracted from the waveforms on the $i$th PMT, where $j \in \mathbb{N}$ and $j < n_i$.  Throughout this paper, we use \textit{PE} to be the short for ``PE counts''. \emph{Detector response model} predicts the PE $\lambda_i$ and timing $T_i$. $\lambda_i$ is the average of the Poissonian $n_i$. $T_i$ is a shift to the $t_{ij}$'s \emph{probability distribution function}~(PDF) $R^0(t - T_i)$, which is determined by $t_0$, TOF, TT and $\psi(t)$.

\subsection{PE and timing prediction}
\label{subsec: PE and timing prediction}
$\lambda_i$ and $T_i$ are the functions of $(\mathbf{r},\mathbf{r}_\mathrm{PMT})$. If the detector is spherically symmetric, the function can be rewritten in $(r, \theta)$. Here $r$ is the vertex radius and $\theta$ is the central angle defined by PMT and vertex. By \emph{varying coefficient model}~\cite{hastie01statisticallearning}, we fit the conditional distribution of $\theta$ given $r$, then fit the coefficients with $r$ by the method of least squares.

In the first step, $n_i$ is modeled by Poisson regression~\cite{10.2307/2344614} as count data.  $\lambda_i$ is connected to the observation $n_i$ by the Poisson log-likelihood
\begin{equation}
	\begin{aligned}
		\label{eq: LikelihoodPE}
		\log\mathscr{L} = -\sum_i \lambda_i + \sum_i n_i\log\lambda_i + \mathrm{Const.}
	\end{aligned}
\end{equation}
A logarithm, called a \emph{link function} in generalized linear model~\cite{10.2307/2344614} terminology, connects $\lambda_i$ with the predictor variable as a linear combination of Legendre polynomials $P_l(\cos\theta)$. $l$ is the order and $c^\lambda_l(r)$'s are the regression coefficients.
\begin{equation}
	\label{eq: linkPE}
	\lambda_i(r, \theta_i) = \mathbb{E}\left[n_i \mid r, \theta_i\right] = \exp\left\{\sum_l c^\lambda_l(r) P_l(\cos\theta_i) + \log E + \mu_i\right\}
\end{equation}
Here $E$ is the energy, $\mu_i$ is the correction term to match the PMT-specific differences dominated by QE. We restrict $\sum_i \mu_i = 0$, $E$ and $\mu_i$ enter the regression as offsets.  Eqs.~\eqref{eq: LikelihoodPE} and \eqref{eq: linkPE} models the dependence on $\theta_i$ and outputs a set of $r$-dependent coefficients $c^\lambda_l(r)$.

Similarly, quantile regression~\cite{davino2013quantile} is used to model timing, reducing the influence of the heavy tail of $R^0(t - T_i)$. The predicted $T_i$ for the $\tau$-quantile is
\begin{equation}
	\label{eq: link_time}
	T_i(r, \theta_i) = \arg\underset{c}{\min} \left[\sum_{j}\mathscr{R}_\tau (t_{ij}-c)\right] = \sum_l c^T_l(r) P_l(\cos\theta_i) + U_i + t_0,
\end{equation}
where $\mathscr{R}_\tau(t)$ is the loss function defined by
\begin{equation}
	\label{eq: link_loss}
	\mathscr{R}_\tau(t) = \left\{
	\begin{array}{ll}
		- t(1-\tau) & \mbox{, if $t < 0$;}    \\
		t \tau      & \mbox{, if $t \geq 0$.}
	\end{array}
	\right.
\end{equation}
$t_0$ is the time offset and $U_i$ is the PMT-specific offset dominated by TT with $\sum_i U_i = 0$. Like $c^\lambda_l(r)$, $c^T_l(r)$ encodes the $\theta$ dependence of timing.

\begin{figure}[!htbp]
	\centering
	\begin{minipage}[htb]{0.45\textwidth}
		\centering
		\includegraphics[width=\textwidth, page=1]{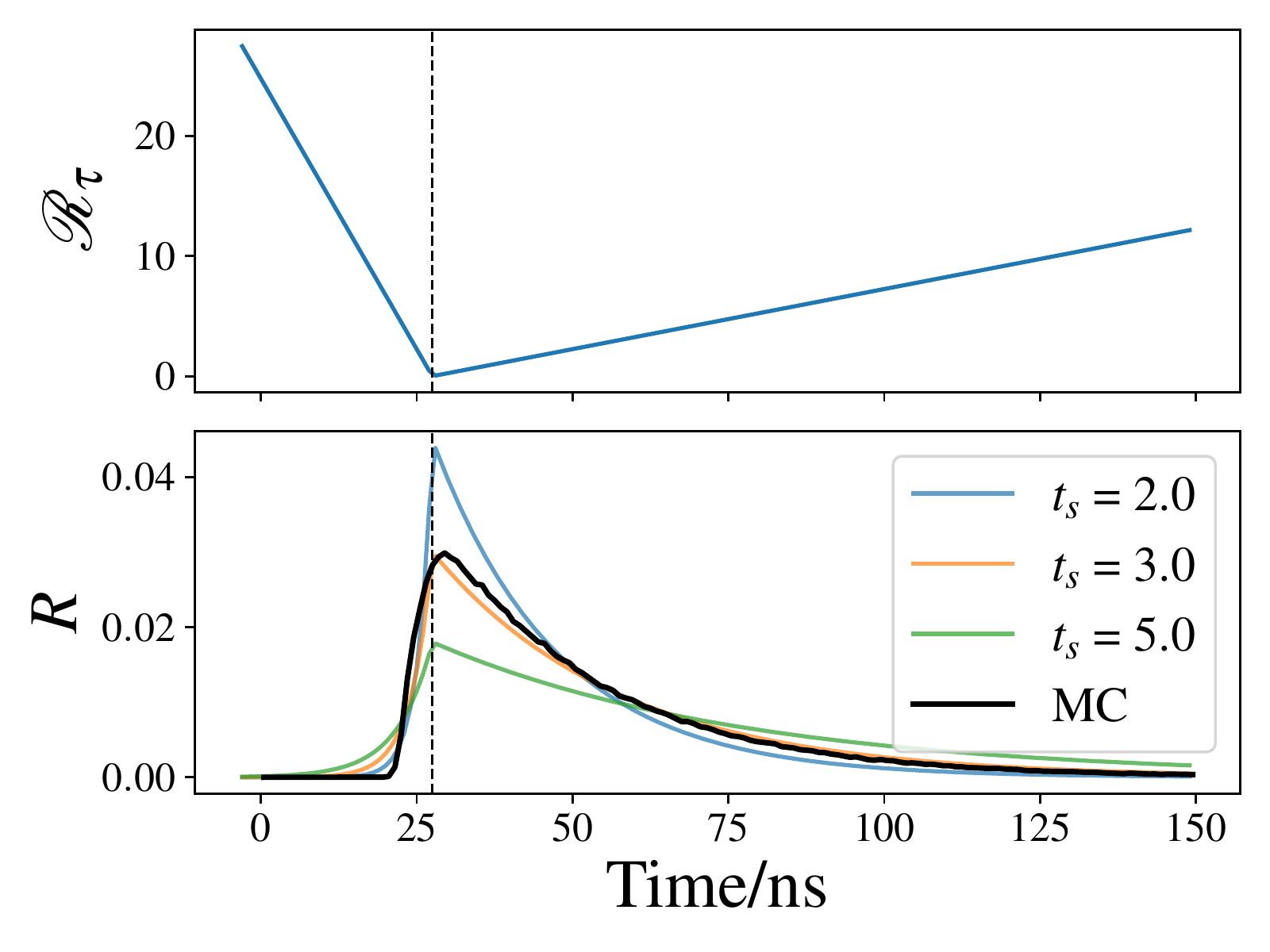}
		\subcaption{The shape of $\mathscr{R}_{\tau}$(top) and $R$(bottom)}
		\label{fig: QRpdf}
	\end{minipage}
	\quad
	\begin{minipage}[htb]{0.45\textwidth}
		\centering
		\includegraphics[width=\textwidth, page=2]{QRpdf.pdf}
		\subcaption{The CDF of $R(t-T_i)$}
		\label{fig: QRcdf}
	\end{minipage}
	\caption{(\subref{fig: QRpdf}) shows the $\mathscr{R}_{\tau}(t-T_i)$ (top) and $R(t-T_i)$ (bottom). $T_i$ is \SI{27.5}{ns} and $\tau$ is 0.1. ``MC'' represents \(R^0(t)\). It is the histogram of simulated event timings at the detector center, see section~\ref{subsec: The training and validation sets}. The best $t_\mathrm{s}$ is \SI{3.0}{ns}. (\subref{fig: QRcdf}) is the \emph{cululative distribution function}~(CDF) of \(R^0(t)\) as ``MC'' and $R(t-T_i)$ as ``Fit''. The steepest slope is at the peak of \(R^0(t)\), leading to $\tau \approx 0.1$.}
\end{figure}

Minimizing \(\mathscr{R}_\tau(t)\) in eqs.~\eqref{eq: link_time} and \eqref{eq: link_loss} is equivalent to maximizing a likelihood function $\mathscr{L} \propto \exp(-\mathscr{R}_{\tau}/t_\mathrm{s})$, where $t_\mathrm{s}$ is an arbitrary positive real number encoding a \emph{time scale}. The normalizing constant is
\begin{equation}
	\begin{aligned}
		\label{eq: Integrate}
		\int_{-\infty}^{\infty} \exp(-\frac{\mathscr{R}_{\tau}(t)}{t_\mathrm{s}}) \mathrm{d} t = \frac{t_\mathrm{s}}{1-\tau} + \frac{t_\mathrm{s}}{\tau} = \frac{t_\mathrm{s}}{\tau(1-\tau)}
	\end{aligned}
\end{equation}
Therefore, quantile regression approximates the timing PDF $R^0(t)$ by
\begin{equation}
	\label{eq: quantilepdf}
	R(t) = \frac{\tau(1-\tau)}{t_\mathrm{s}}\exp(-\frac{\mathscr{R}_{\tau}(t)}{t_\mathrm{s}}).
\end{equation}
The examples of the loss function $\mathscr{R}_\tau(t)$ and timing PDF $R(t)$ are shown in figure~\ref{fig: QRpdf}.

In the second step, we use another set of Legendre polynomials to fit coefficients $c^\lambda_l(r)$ and $c^T_l(r)$. $r$ is scaled to $[-1,1]$ by dividing the LS radius $r_\mathrm{LS}$. Due to the symmetry, we only use the even orders to guarantee the $c^\lambda_l(r)$ and $c^T_l(r)$'s derivatives are 0 at the detector center,
\begin{equation}
	\label{eq: fit_r_PE}
	c^\lambda_l(r) = \sum_{m=0}^{\infty} {c^\lambda_{l,2m}} P_{2m}(r/r_\mathrm{LS}), \quad \quad c^T_l(r) = \sum_{m=0}^{\infty} {c^T_{l,2m}} P_{2m}(r/r_\mathrm{LS}).
\end{equation}

The above varying coefficient model requires simulation or calibration at fixed radii.  Alternatively, if the simulated events are uniformly distributed in the detector, the above 2-step requirement can be relaxed with a one-step regression,
\begin{equation}
	\label{eq: fit2d_PE}
	\lambda_i(r, \theta_i) = \exp\left[\sum_m c^\lambda_{m} F_m(r, \cos\theta_i)\right], \quad \quad T_i(r,\theta_i) = \sum_m c^T_{m} F_m(r, \cos\theta_i).
\end{equation}
$F_m(r, \cos\theta)$ is a binary basis function of $r$ and $\theta$ at the $m$th order such as Zernike polynomials~\cite{noll1976zernike}. Another way to construct a binary basis function is to product the two Legendre polynomials from the varying coefficient model, called \emph{double Legendre}.
\begin{equation}
	F_{l,m}(r,\cos\theta) = P_m(r/r_\mathrm{LS}) \times P_l(\cos\theta)
\end{equation}
which is indexed by two subscripts $l$ and $m$. Due to the memory constraints of our computing system, a regression cannot handle more than 800 parameters in one pass.  Consequently binary basis models are more restricted than the varying coefficient one, although the former models more symmetrically handle $r$ and $\theta$.  Their best order selections are discussed in section~\ref{subsec: The training and validation sets}.

\subsection{Training and validation}
\label{subsec: The training and validation sets}
\begin{figure}[!htbp]
	\centering
	\includegraphics[width=0.75\linewidth,clip]{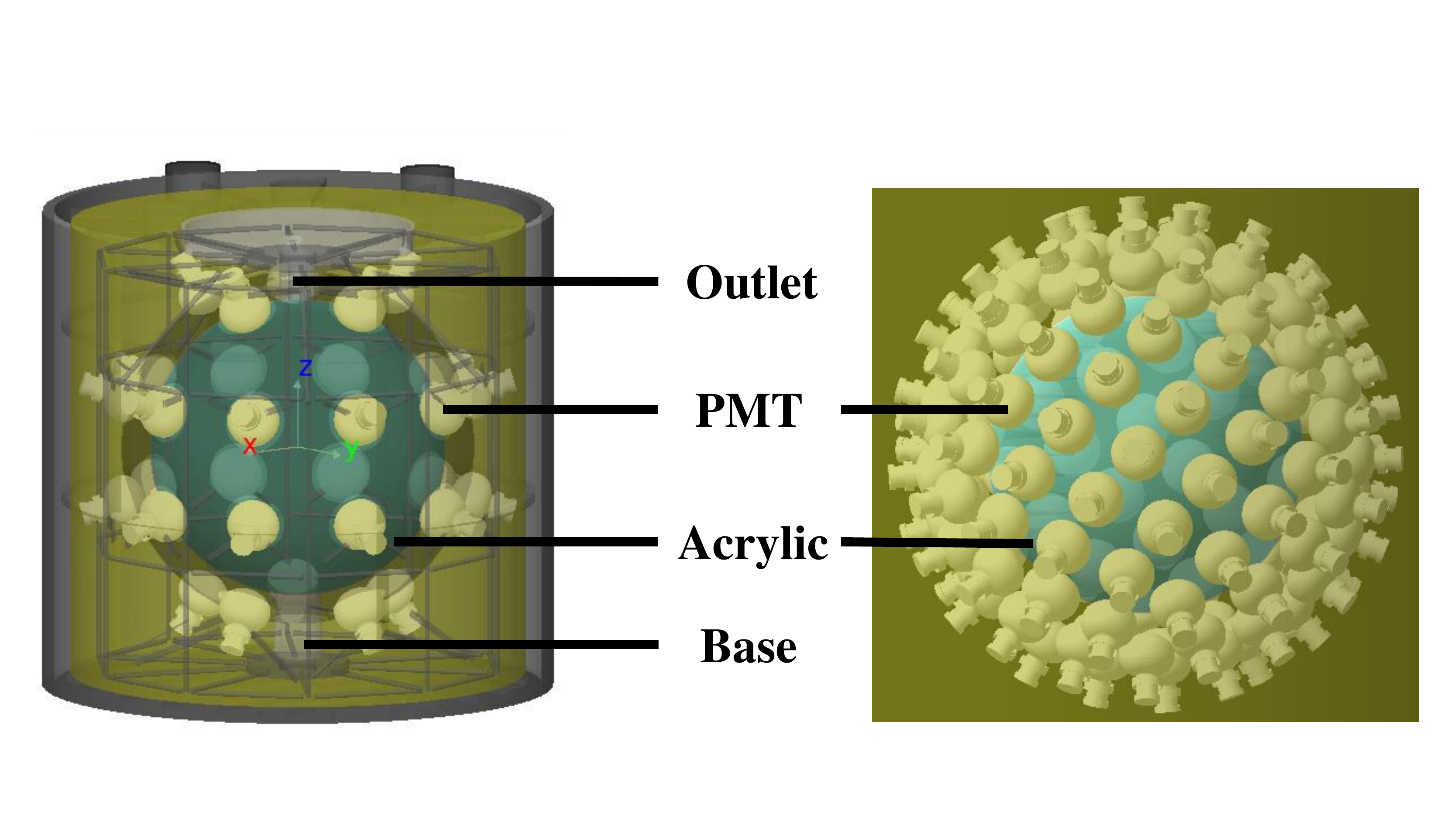}
	\caption{Geometries of the Jinping prototype (left) and the ideal detector (right).}
	\label{fig: Jinping}
\end{figure}

The training dataset is generated by a custom piece of software, \emph{Jinping Simulation and Analysis Package}~(JSAP), based on GEANT4~\cite{agostinelli2003geant4}. Two geometries are defined as in figure~\ref{fig: Jinping}: the first utilizes the \emph{Jinping prototype}~\cite{WANG201781}, featuring 30 PMTs, and the second is an \emph{ideal detector} upgraded to 120 PMTs, following the Fibonacci arrangement~\cite{gonzalez2010measurement}. The number of PMTs in the ideal detector follows the criterion in section~\ref{subsec: criterion}. The central LS~\cite{guo2019slow} is in an acrylic shell with a water buffer. The Jinping prototype has an outlet and a support base, which are removed in the ideal detector. Table~\ref{tab: par} lists some essential parameters. In reality, the QE and TT for each PMT are different, and the TTS affects $R^0(t)$. We ignore such difference since $\psi(t)$ dominates the $R^0(t)$. $\mu_i$ in eq. \eqref{eq: LikelihoodPE} and $U_i$ in eq. \eqref{eq: link_time} can be obtained by calibration. Without loss of generality, we assume PMTs are identical by setting $\mu_i = U_i = 0$.
\begin{table}[!htbp]
	\centering
	\caption{Important parameters in simulation}
	\begin{tabular}{ccc|cc}
		\toprule
		parameter        & value (Jinping prototype) & value (ideal detector) & parameter & value        \\
		\midrule
		$r_\mathrm{LS}$  & \SI{650}{mm}              & \SI{650}{mm}           & QE        & 0.2          \\
		$r_\mathrm{PMT}$ & \SI{832}{mm}              & \SI{900}{mm}           & TTS       & \SI{2.2}{ns} \\
		number of PMTs   & 30                        & 120                    & $t_0$     & \SI{0}{ns}   \\
		\bottomrule
	\end{tabular}
	\label{tab: par}
\end{table}

$\psi(t)$ is parameterized in eq. \eqref{eq: TimeProfile} by the \emph{decay time constant} $\tau_d=$ \SI{26.0}{ns} and the \emph{rise time constant}  $\tau_r=$ \SI{1.6}{ns},
\begin{equation}
	\label{eq: TimeProfile}
	\mathrm{\psi}(t) \sim \exp\left(-\frac{t}{\tau_d}\right)\left[1-\exp\left(-\frac{t}{\tau_r}\right)\right].
\end{equation}
The average refraction index of LS is 1.48, causing TR to occur at $r>0.88 r_\mathrm{LS}$ which is \SI{570}{mm} for the Jinping prototype.

\begin{figure}[!htbp]
	\centering
	\begin{minipage}[htb]{\textwidth}
		\centering
		\includegraphics[width=\linewidth, trim=25 47 0 0, clip]{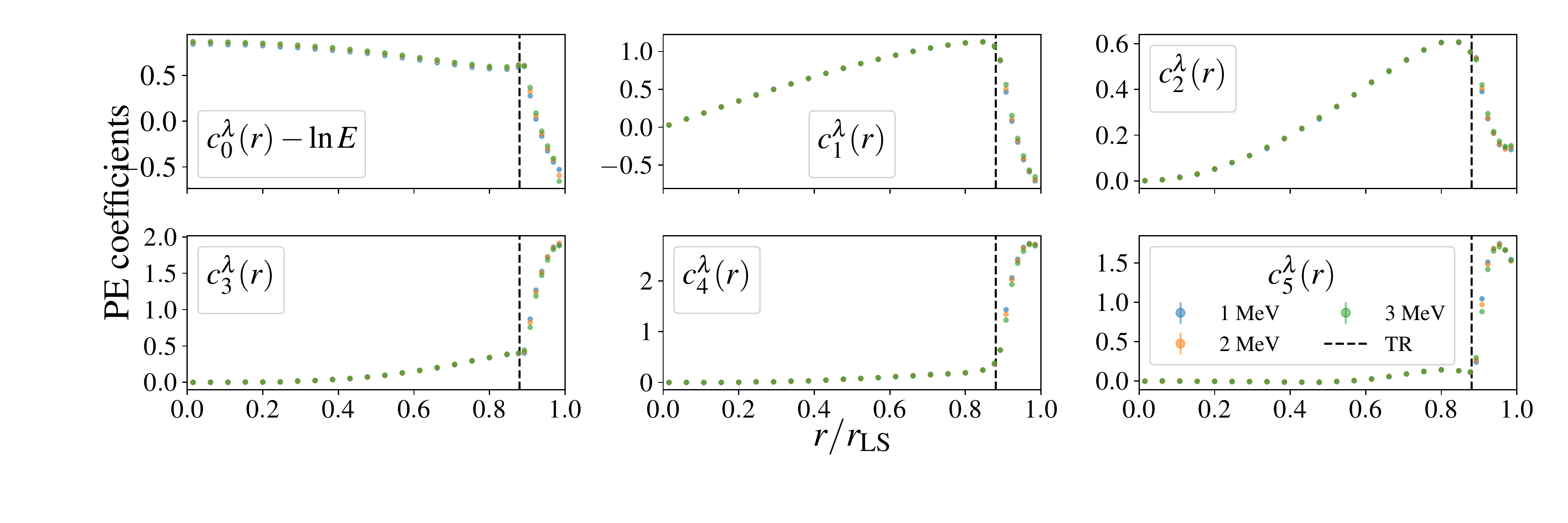}
		\subcaption{First six $c^\lambda_l(r)$ of Jinping prototype under different energies.}
		\label{fig: PECoeff}
	\end{minipage}
	\hfill
	\begin{minipage}[htb]{\textwidth}
		\centering
		\includegraphics[width=\linewidth, trim=25 47 0 0, clip]{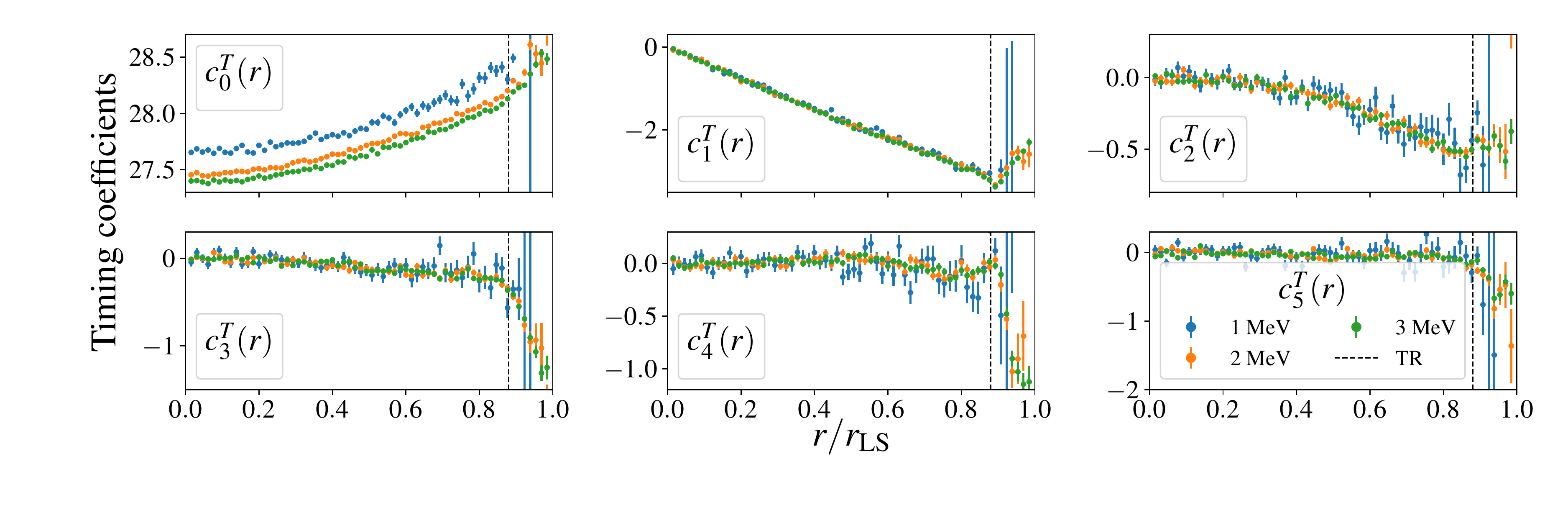}
		\subcaption{First six $c^T_l(r)$ of Jinping prototype under different energies.}
		\label{fig: TimeCoeff}
	\end{minipage}
	\hfill
	\centering
	\begin{minipage}[htb]{0.39\textwidth}
		\centering
		\includegraphics[width=\linewidth, trim=0 10 25 0, clip, page=1]{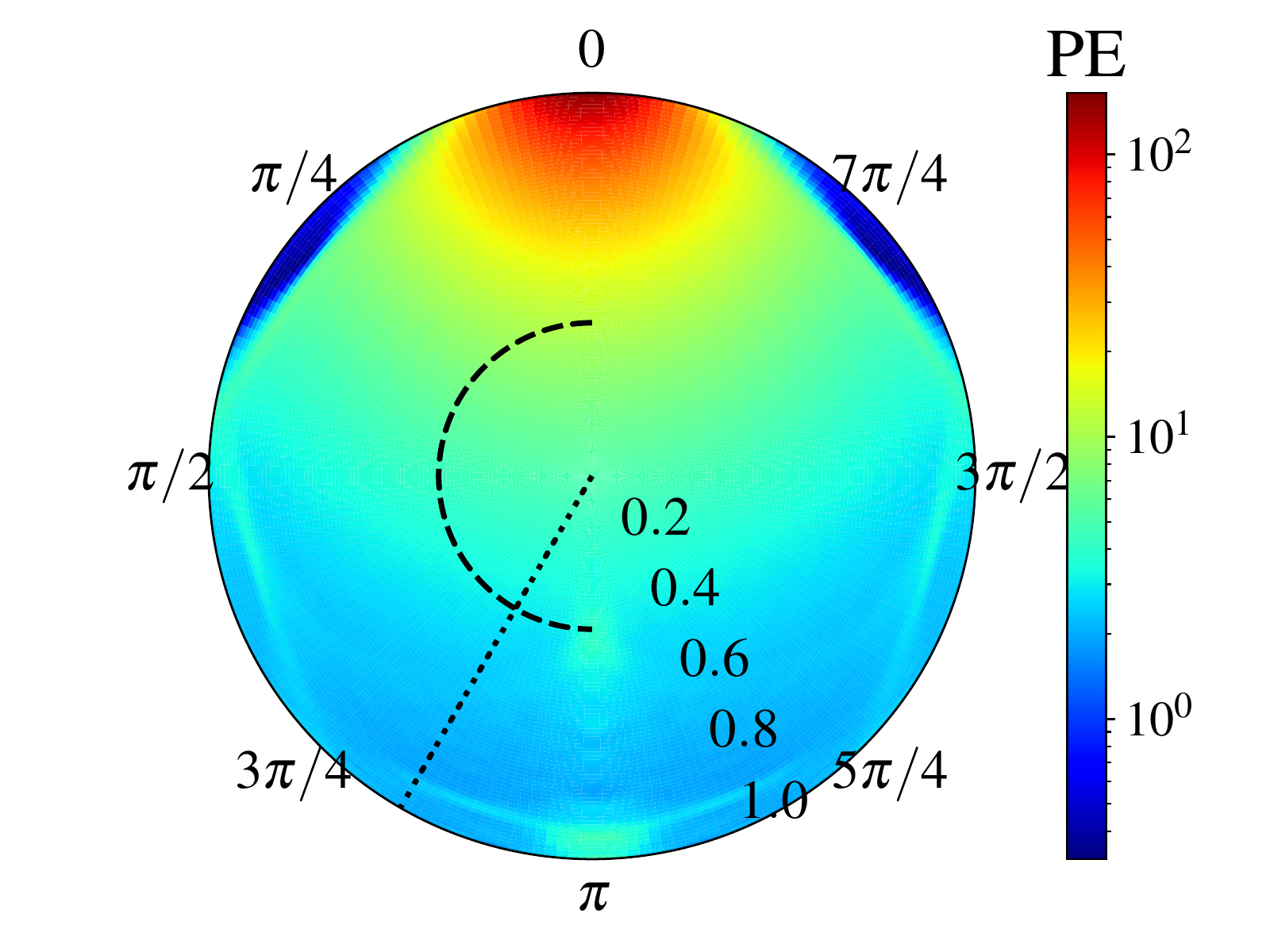}
		\subcaption{Predicted PE}
		\label{fig: Probe_PE}
	\end{minipage}
	\quad
	\begin{minipage}[htb]{0.39\textwidth}
		\centering
		\includegraphics[width=\linewidth, trim=0 10 25 0, clip, page=4]{probe.pdf}
		\subcaption{Ratio of simulation over predicted PEs}
		\label{fig: quotient}
	\end{minipage}
	\hfill
	\begin{minipage}[htb]{0.39\textwidth}
		\centering
		\includegraphics[width=\linewidth, trim=0 10 25 0, clip, page=2]{probe.pdf}
		\subcaption{Predicted timing}
		\label{fig: Probe_Time}
	\end{minipage}
	\quad
	\begin{minipage}[htb]{0.39\textwidth}
		\centering
		\includegraphics[width=\linewidth, trim=0 10 0 0, clip, page=5]{probe.pdf}
		\subcaption{Inhomogeneous Poisson process}
		\label{fig: Poisson_flux}
	\end{minipage}
	\caption{(\subref{fig: PECoeff}) and (\subref{fig: TimeCoeff}) are regression results. TR occurs at the $0.88 r_{\mathrm{LS}}$ (dashed line). The error bars are $\pm 1$ standard deviations. (\subref{fig: Probe_PE}) -- (\subref{fig: Poisson_flux}) uses \SI{2}{MeV} $e^{-}$ in Jinping prototype. (\subref{fig: Probe_PE}) and (\subref{fig: Probe_Time}) are the PE and timing responses, respectively. In (\subref{fig: Probe_PE}), we shall look into the focus structures in the slices of the dashed line $r=0.4r_\mathrm{LS}$ and dotted line $\theta=5\pi/6$ in section~\ref{subsec: Other reflections by acrylic shell}. (\subref{fig: quotient}) is the ratio between the predicted PE and its MC truth. (\subref{fig: Poisson_flux}) shows the average function of the inhomogeneous Poisson process compared with the MC at $r = 0.99 r_\mathrm{LS}$ and $\theta = 0$.}
	\label{fig: probe}
\end{figure}

We focus on $e^{-}$ in this work. The training dataset includes multiple batches. Vertices in each batch have the same $r$, ranging from \SI{0}{mm} to \SI{550}{mm} with step \SI{10}{mm}, and from \SI{550}{mm} to \SI{640}{mm} with step \SI{2}{mm}. Figure~\ref{fig: PECoeff} subtracts $\log E$ in the 0th order and the first six orders of $c^\lambda_l$ under 1, 2 and 3 \si{MeV}. Figure~\ref{fig: TimeCoeff} shows the first six orders of $c^T_l$. The coefficient changes dramatically at $0.88r_{\mathrm{LS}}$, where the TR just happens.  It is evident from figure~\ref{fig: PECoeff} and \ref{fig: TimeCoeff} that \(c^\lambda_l\) and \(c^T_l\) depend on energy only when $l=0$.  Therefore one energy sample alone captures the detector response in any other energies. The Jinping prototype utilizes a 25-PMT threshold~\cite{zhao2022measurement} and the trigger efficiency drops for events near the boundary.  We take \SI{2}{MeV} events that are energetic enough to guarantee \SI{\sim 100}{\percent} trigger efficiency.

We can express $\lambda_i(r, \theta_i)$ and $T_i(r,\theta_i)$ in a single heat map of a disk. Figure~\ref{fig: Probe_PE} and \ref{fig: Probe_Time} shows the predicted PE and timing, respectively. The ratio of the predicted PE and its truth is almost 1 in figure~\ref{fig: quotient}, showing a good fit.  The models of PE and timing at a specific $r,\theta$ can be combined into an inhomogeneous Poisson process with average function $\lambda_i(r, \theta_i) R(t-T_i(r, \theta_i))$, shown in figure~\ref{fig: Poisson_flux}.

\begin{figure}[!htbp]
	\begin{minipage}[htb]{0.45\textwidth}
		\centering
		\includegraphics[width=\linewidth, page=2]{score.pdf}
		\subcaption{Relative scores for different sub-models}
		\label{fig: validate}
	\end{minipage}
	\quad
	\begin{minipage}[htb]{0.45\textwidth}
		\centering
		\includegraphics[width=\linewidth, page=1]{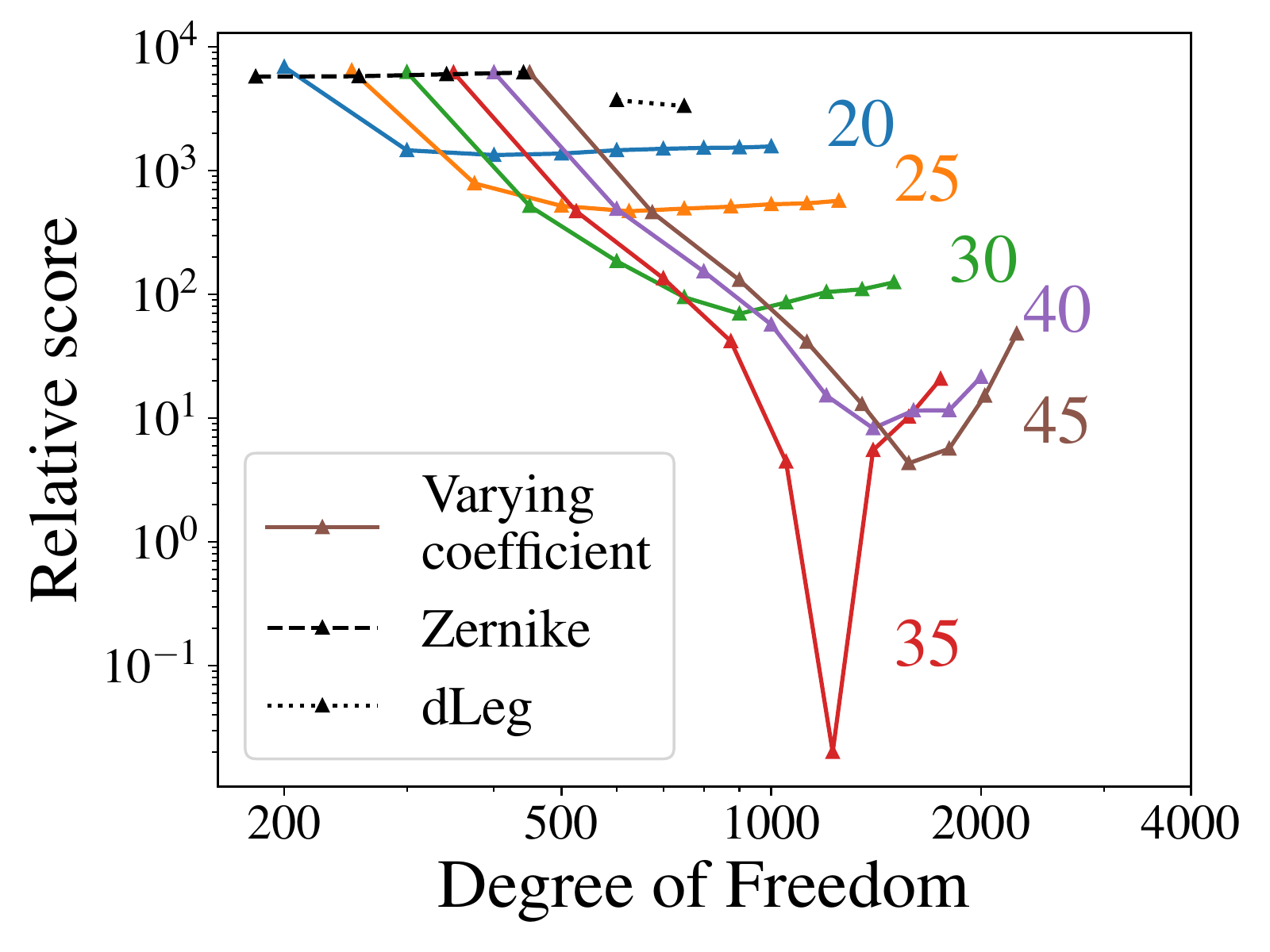}
		\subcaption{Relative scores vs the degree of freedom}
		\label{fig: validate_all}
	\end{minipage}
	\caption{Relative scores to the best model. (\subref{fig: validate}) indicates the number of parameters $35\times35$ of $(\theta, r)$ to be the best. (\subref{fig: validate_all}) adds the binary polynomials. The numbers are the number of parameters on $\theta$ in the varying coefficient models. ``\emph{Zernike}'' and ``\emph{dLeg}'' are Zernike and double Legendre polynomials, respectively.}

\end{figure}

To determine the optimal number of parameters, we simulate another 15000 events as the validation dataset.  A better goodness of fit for PE is manifested by a higher score in log-likelihood given by eqs.~\eqref{eq: LikelihoodPE} and \eqref{eq: fit_r_PE}.  Figure~\ref{fig: validate} shows the highest score requires the number of parameters to be $35\times35$ of $(\theta, r)$. The binary polynomials are shown for comparison in figure~\ref{fig: validate_all}. The number of parameters is up to 441 for Zernike and 750 for double Legendre polynomials. Their scores are limited by the number of parameters.

It is difficult to make a similar selection for $c^T_{l,m}$ because their scores rely on $\tau$ and $t_\mathrm{s}$. Since PE is dominant in the small detectors~\cite{GALBIATI2006700}, we choose $10\times35$ on $(\theta, r)$ for $c^T_{l,m}$ to balance speed and accuracy.

\subsection{Likelihood for reconstruction}
\label{subsec: Likelihood for reconstruction}
We use MLE for reconstruction with the likelihood function satisfying
\begin{equation}
	\begin{aligned}
		\log\mathscr{L}(\mathbf{r}, E, t_0) & \sim \overbrace{\sum_i\sum_j \frac{\mathscr{R}_\tau(t_{ij} - T_i - t_0)}{t_\mathrm{s}}}^{\mathrm{timing\ part}} + \overbrace{\sum_i \left(-\frac{E\lambda_i}{E_0} +  n_i\log \frac{E\lambda_i}{E_0} \right)}^{\mathrm{PE\ part}}.
		\label{eq: MC_Likelihood}
	\end{aligned}
\end{equation}
The former is the timing part and the latter is the PE part. $E$ only contributes to the PE part. $E_0$ is energy (\SI{2}{MeV}) in the training dataset. The unbiased energy estimation of eq. \eqref{eq: MC_Likelihood} is
\begin{equation}
	\begin{aligned}
		\frac{\partial \log \mathscr{L}}{\partial E} = 0 \Rightarrow \hat{E} = \frac{\sum_i n_i}{\sum_i {\lambda}_i(\mathbf{r})} E_0
	\end{aligned}
	\label{eq: Fix}
\end{equation}
We use \emph{Sequential Least Squares Programming}~\cite{kraft1988software} to maximize eq.~\eqref{eq: MC_Likelihood}. The energy is calculated in each iteration by eq. \eqref{eq: Fix} to reduce the time costs.

Most gradient-based methods are local optimizers. The origin of the local maxima will be discussed in section~\ref{sec3}. Our strategy to obtain the global maximum is as follows. We generate two grids to calculate the predicted PE. Each grid is $30\times 50\times 50$ equally spaced on $r$, $\cos\theta_\mathrm{v}$ and $\phi_\mathrm{v}$. The inner grid covers the radius of \SIrange{0}{570}{mm} and the outer covers the rest. We choose the best points of the inner and outer grids as initial values for the gradient optimizer. The larger of the two $\log\mathscr{L}$'s is recorded.

\section{Multimodality of the likelihood function}
\label{sec3}

The convexity of the minus-log-likelihood in eq.~\eqref{eq: MC_Likelihood} is crucial for reconstruction. The reconstruction results are sensitive to the initial values of the gradient optimizer if the minus-log-likelihood is not convex, or equivalently, the likelihood function is \emph{multimodal}. Using a poor vertex undermines the energy resolution and the background rejection by fiducial volume cuts.

In this section, we concentrate on PE~(but not timing) since it is dominant in the small detectors~\cite{GALBIATI2006700}. We replace the likelihood function in eq.~\eqref{eq: MC_Likelihood} with \emph{cosine distance} to eliminate the influence of fluctuations in the observed PEs. We further narrow our focus to the 3 closest PMTs, giving a criterion of multimodality in the likelihood function.

\subsection{Cosine distance}
\label{subsec: cosine distance}
We define \emph{pattern} $\Lambda$ as a vector containing the predicted PE on the PMT space.
\begin{equation}
	\Lambda = \{\lambda_1, \lambda_2, \cdots, \lambda_{N_\mathrm{PMT}}\}
\end{equation}
If $\mathbf{r}_1$ and $\mathbf{r}_2$ are both the solution for an event and $\mathbf{r}_1 \neq \mathbf{r}_2$, we have $\Lambda(\mathbf{r}_1) = \alpha\Lambda(\mathbf{r}_2)$, and $\alpha$ is an arbitrary positive number. The \emph{cosine distance}
\begin{equation}
	\label{eq: cosdist}
	d_{\mathrm{cos}}\left[\Lambda(\mathbf{r}_1) \mid \Lambda(\mathbf{r}_2)\right] = 1 - \frac{\Lambda(\mathbf{r}_1)\cdot\Lambda(\mathbf{r}_2)}{\norm{\Lambda(\mathbf{r}_1)}\norm{\Lambda(\mathbf{r}_2)}}
\end{equation}
is zero.

\begin{figure}[!htbp]
	\centering
	\begin{minipage}[htb]{0.31\textwidth}
		\centering
		\includegraphics[width=\linewidth]{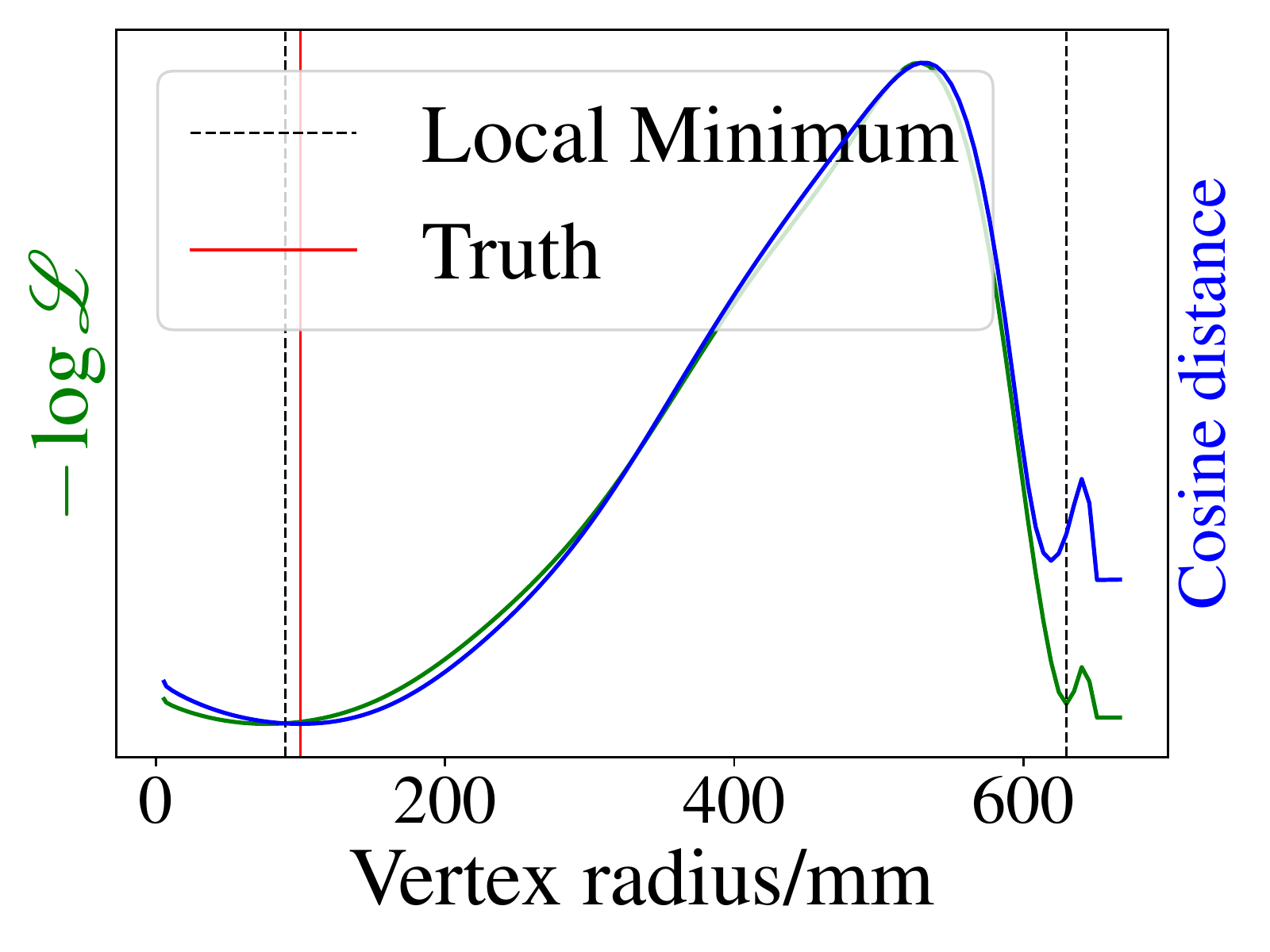}
		\subcaption{Vertex at $(0, 0, 100)$~\si{mm}}
		\label{fig: Likelihood0}
	\end{minipage}
	\quad
	\begin{minipage}[htb]{0.31\textwidth}
		\centering
		\includegraphics[width=\linewidth]{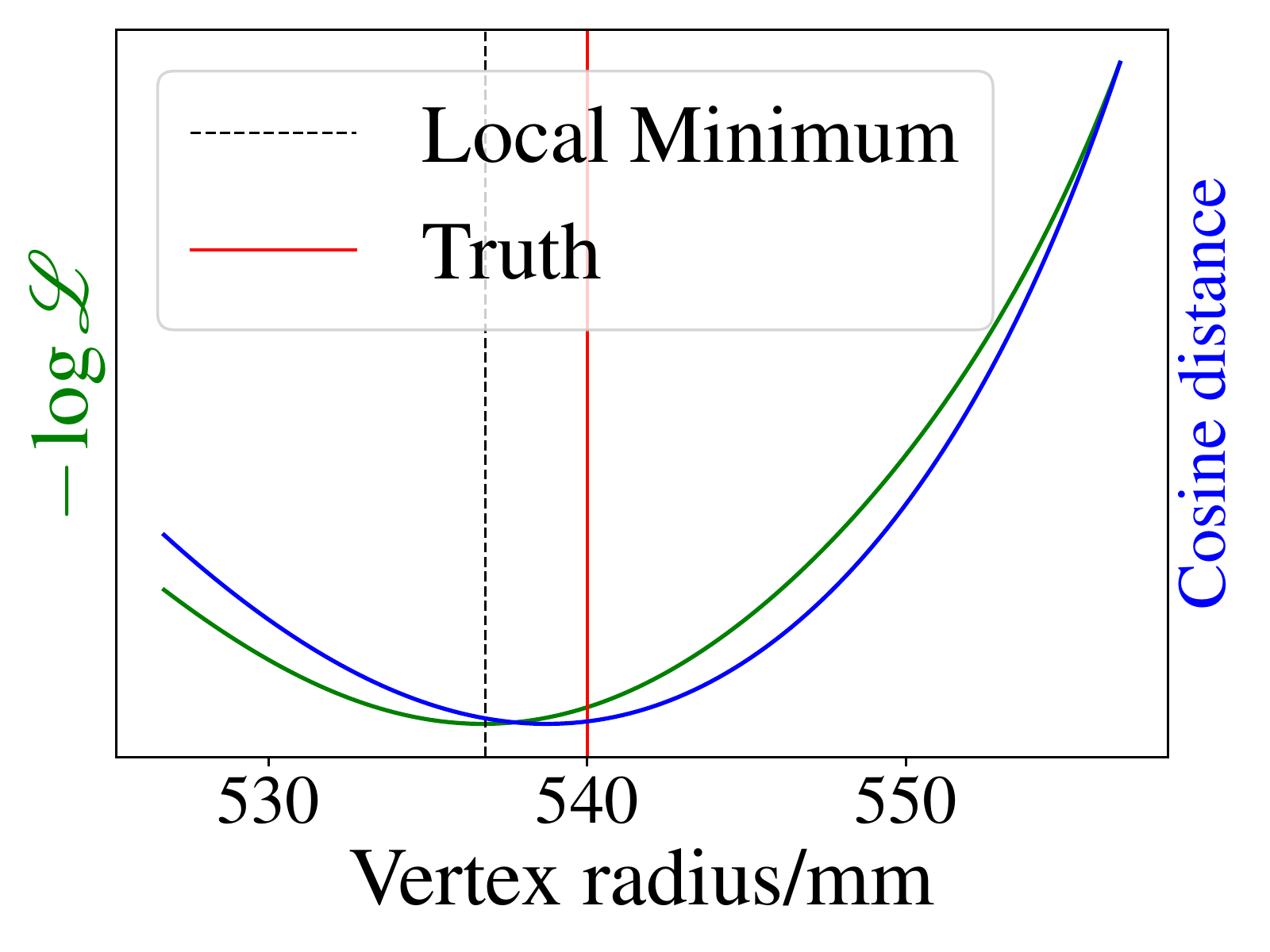}
		\subcaption{vertex at $(540, 0, 0)$~\si{mm}}
		\label{fig: Likelihood1}
	\end{minipage}
	\quad
	\begin{minipage}[htb]{0.31\textwidth}
		\centering
		\includegraphics[width=\linewidth]{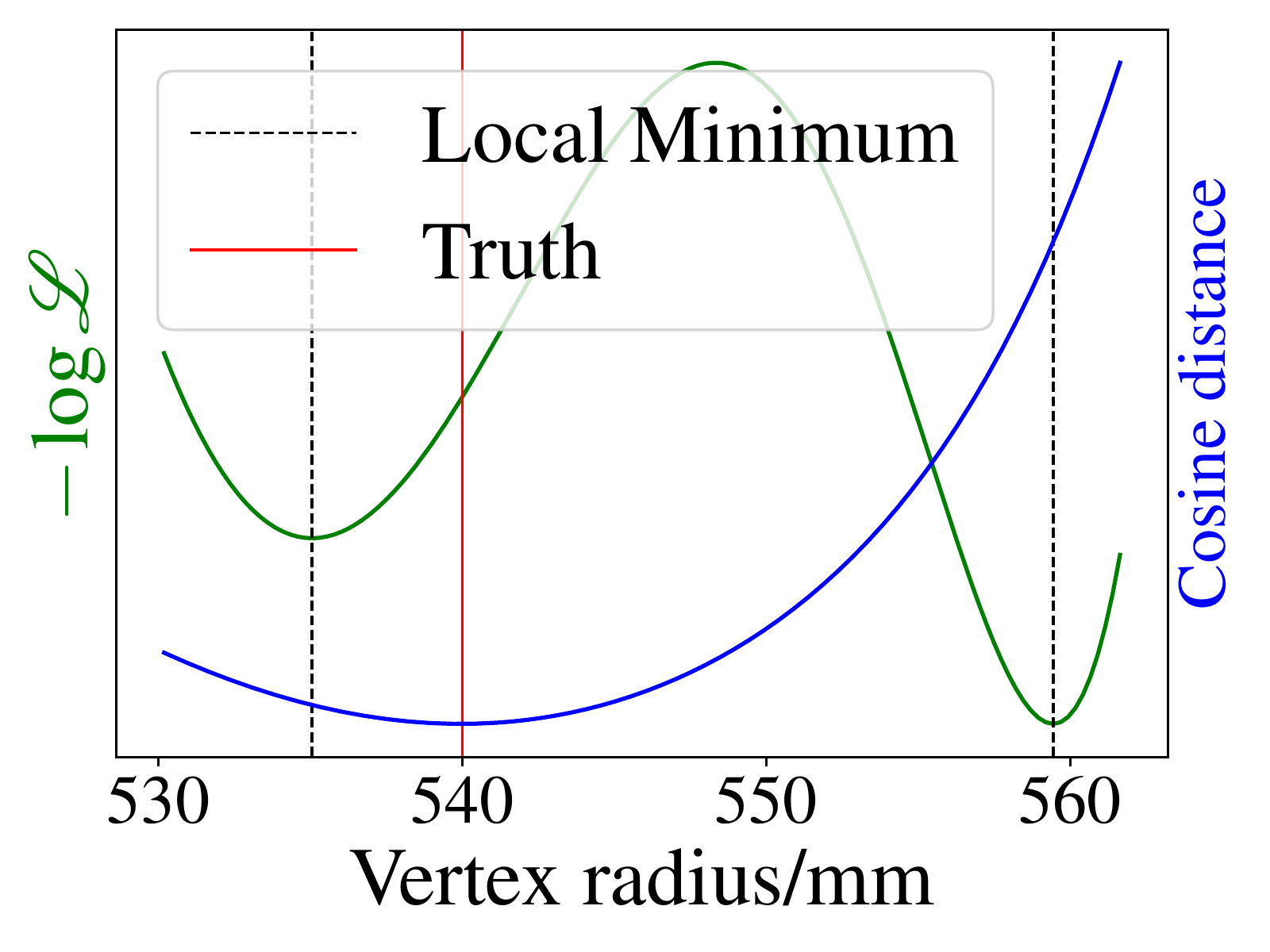}
		\subcaption{Vertex at $(540, 0, 0)$~\si{mm}}
		\label{fig: Likelihood2}
	\end{minipage}
	\caption{Comparison of $d_{\cos}$ with $-\log \mathscr{L}$. All events at $(0, 0, 100)$ \si{mm} degenerate with $r=$~\SI{600}{mm} on the $z$ axis. (\subref{fig: Likelihood0}) is one typical event and the two local solutions are in black dashed lines. The true vertex is in the red line. The green line scans the $-\log\mathscr{L}$ and the blue line scans the $d_{\cos}$, showing that they both have two local minima. (\subref{fig: Likelihood1}) and (\subref{fig: Likelihood2}) use true events at $(540, 0, 0)$ \si{mm}, comparing one unimodal with one multimodal event. $d_{\cos}$ itself is convex. The difference in $-\log\mathscr{L}$ between (\subref{fig: Likelihood1}) and (\subref{fig: Likelihood2}) is from fluctuations.}
\end{figure}

$d_\mathrm{cos}$ matches well with the $-\log \mathscr{L}$. Figure~\ref{fig: Likelihood0} shows an event at $(0, 0, 100)$~\si{mm} of the Jinping prototype. Two local solutions with different initial values are shown in black dashed lines. The green line scans the $-\log \mathscr{L}$ between the two.  The blue line shows the $d_{\cos}$ of these points by the true expected PEs. Figure~\ref{fig: Likelihood1} and \ref{fig: Likelihood2} are at $(540, 0, 0)$~\si{mm} in the ideal detector. Figure~\ref{fig: Likelihood1} demonstrates a good unimodal event, in which both $d_{\cos}$ and $-\log\mathscr{L}$ are convex.  Figure~\ref{fig: Likelihood2} is an example of multimodal event where $-\log\mathscr{L}$ is not convex, but the $d_{\cos}$ is the same as figure~\ref{fig: Likelihood1}. Observation fluctuations lead to the $-\log\mathscr{L}$ difference between the two. Of all events at $(540, 0, 0)$~\si{mm}, the fraction of bad events like figure~\ref{fig: Likelihood2} is less than \SI{1}{\%}.

\begin{figure}[H]
	\centering
	\begin{minipage}[htb]{0.42\textwidth}
		\centering
		\includegraphics[width=\linewidth, page=1, trim=0 30 0 0, clip]{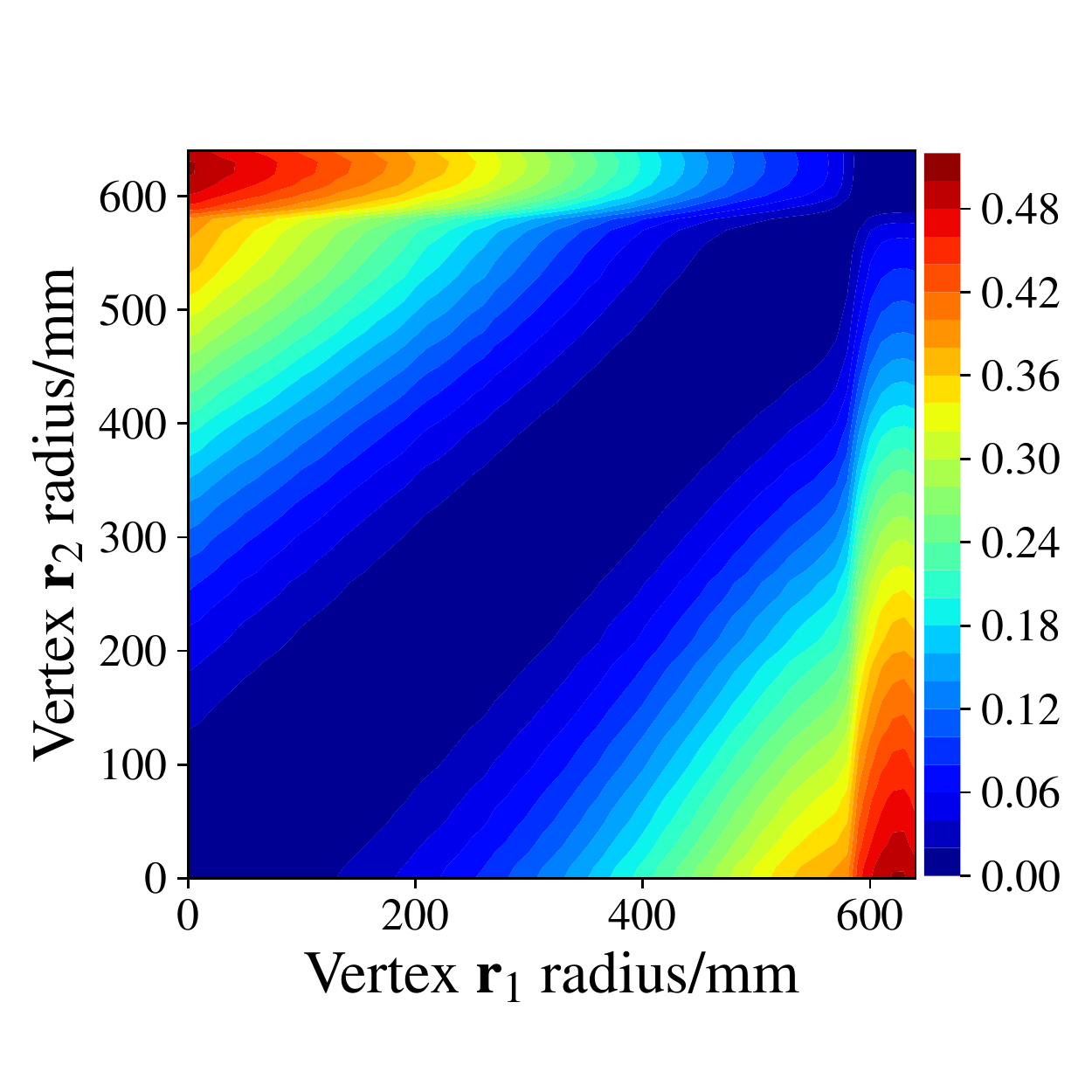}
		\subcaption{cosine distance on $x$ axis}
		\label{fig: cosdist_x}
	\end{minipage}
	\quad
	\begin{minipage}[htb]{0.42\textwidth}
		\centering
		\includegraphics[width=\linewidth, page=2, trim=0 30 0 0, clip]{cosdist.pdf}
		\subcaption{Cosine distance on $z$ axis}
		\label{fig: cosdist_z}
	\end{minipage}
	\hfill
	\begin{minipage}[htb]{0.42\textwidth}
		\centering
		\includegraphics[width=\linewidth, page=1, trim=30 20 0 0]{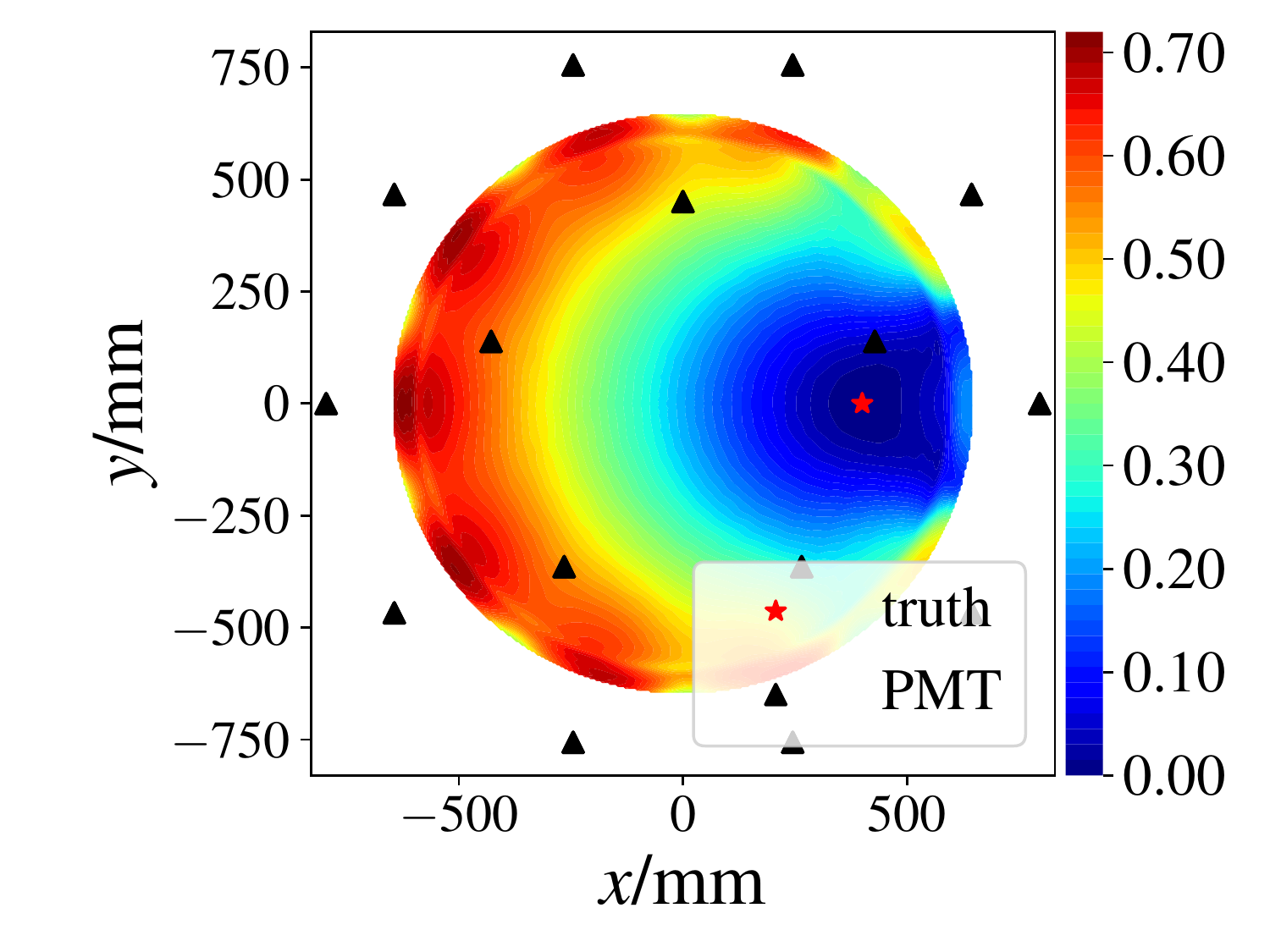}
		\subcaption{Similarity for (400,0,0) \si{mm} to $xOy$ plane}
		\label{fig: plane}
	\end{minipage}
	\quad
	\begin{minipage}[htb]{0.42\textwidth}
		\centering
		\includegraphics[width=\linewidth, page=2, trim=30 20 0 0]{scan_2d.pdf}
		\subcaption{zoom for (\subref{fig: plane})}
		\label{fig: plane_zoom}
	\end{minipage}
	\caption{Cosine distance of different cases. The color shows the value of $d_{\cos}[\Lambda(\mathbf{r}_1) \mid \Lambda(\mathbf{r}_2)]$. (\subref{fig: cosdist_x}) $\mathbf{r}_1$ and $\mathbf{r}_2$ are both on the $x$ axis. (\subref{fig: cosdist_z}) $\mathbf{r}_1$ and $\mathbf{r}_2$ are both on the $z$ axis. The red dashed line shows the center and radius at \SI{600}{mm} degenerate. (\subref{fig: plane}) $\mathbf{r}_1$ is $(400, 0, 0)\ \mathrm{mm}$ and $\mathbf{r}_2$ scans the $xOy$ plane. (\subref{fig: plane_zoom}) is the zoom for (\subref{fig: plane}). The red star and the red circles degenerate.}
	\label{fig: cosdist}
\end{figure}

Figure~\ref{fig: cosdist} maps out multimodality of $d_{\cos}$. $\mathbf{r}_1$ and $\mathbf{r}_2$ are all on the $x$ axis in figure~\ref{fig: cosdist_x}, and on the $z$ axis in figure~\ref{fig: cosdist_z}. The slice in figure~\ref{fig: cosdist_z} shows $z=$~\SI{0}{mm} and \SI{600}{mm} degenerate. In figure~\ref{fig: plane}, $\mathbf{r}_1$ is \SI{400}{mm} on the $x$ axis and $\mathbf{r}_2$ scans the $xOy$ plane. Figure~\ref{fig: plane_zoom} zooms around the true position.  Local minima are found near the detector boundary. The shape of $d_{\cos}$ map is remarkably consistent with the reconstruction results in figure~\ref{fig: Recon}, which will be discussed in section~\ref{subsec: MC data results}.

\subsection{Cosine distance for 3-PMT case}
\label{subsec: Cosine distance for 3-PMT case}

\begin{figure}[H]
	\centering
	\begin{minipage}[htb]{0.31\textwidth}
		\centering
		\includegraphics[trim=45 0 50 0, clip, width=\linewidth, page=1]{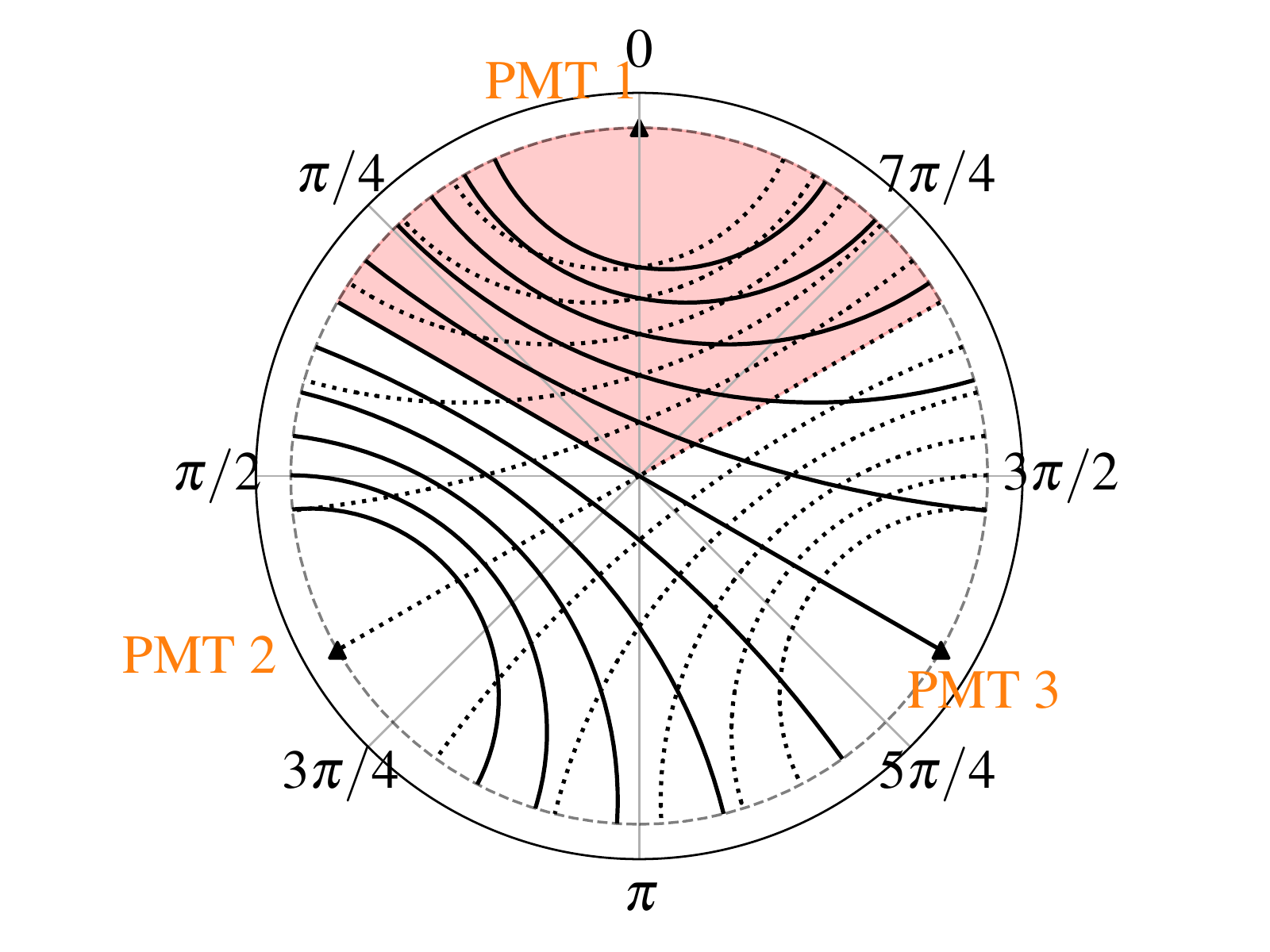}
		\subcaption{ER lines of eq. \eqref{eq: d2}}
		\label{fig: d2}
	\end{minipage}
	%\quad
	\begin{minipage}[htb]{0.31\textwidth}
		\centering
		\includegraphics[trim=45 0 50 0, clip, width=\linewidth, page=2]{equiv.pdf}
		\subcaption{ER lines of eq. \eqref{eq: d2cos}}
		\label{fig: d2cos}
	\end{minipage}
	%\quad
	\begin{minipage}[htb]{0.36\textwidth}
		\centering
		\vspace{1.2em}
		\includegraphics[trim=0 10 0 0, clip, width=\linewidth, page=4]{equiv.pdf}
		\subcaption{ER lines of eq. \eqref{eq: d2cos}}
		\label{fig: ER_100}
	\end{minipage}
	\centering
	\begin{minipage}[htb]{0.31\textwidth}
		\centering
		\includegraphics[trim=40 0 60 0, clip, width=\linewidth, page=7]{equiv.pdf}
		\subcaption{Jinping outlet arrangement}
		\label{fig: Equiv_outlet}
	\end{minipage}
	\quad
	\begin{minipage}[htb]{0.31\textwidth}
		\centering
		\includegraphics[trim=40 0 60 0, clip, width=\linewidth, page=8]{equiv.pdf}
		\subcaption{Jinping general arrangement}
		\label{fig: Equiv_general}
	\end{minipage}
	\quad
	\begin{minipage}[htb]{0.315\textwidth}
		\centering
		\vspace{0.2em}
		\includegraphics[trim=40 0 45 0, clip, width=\linewidth]{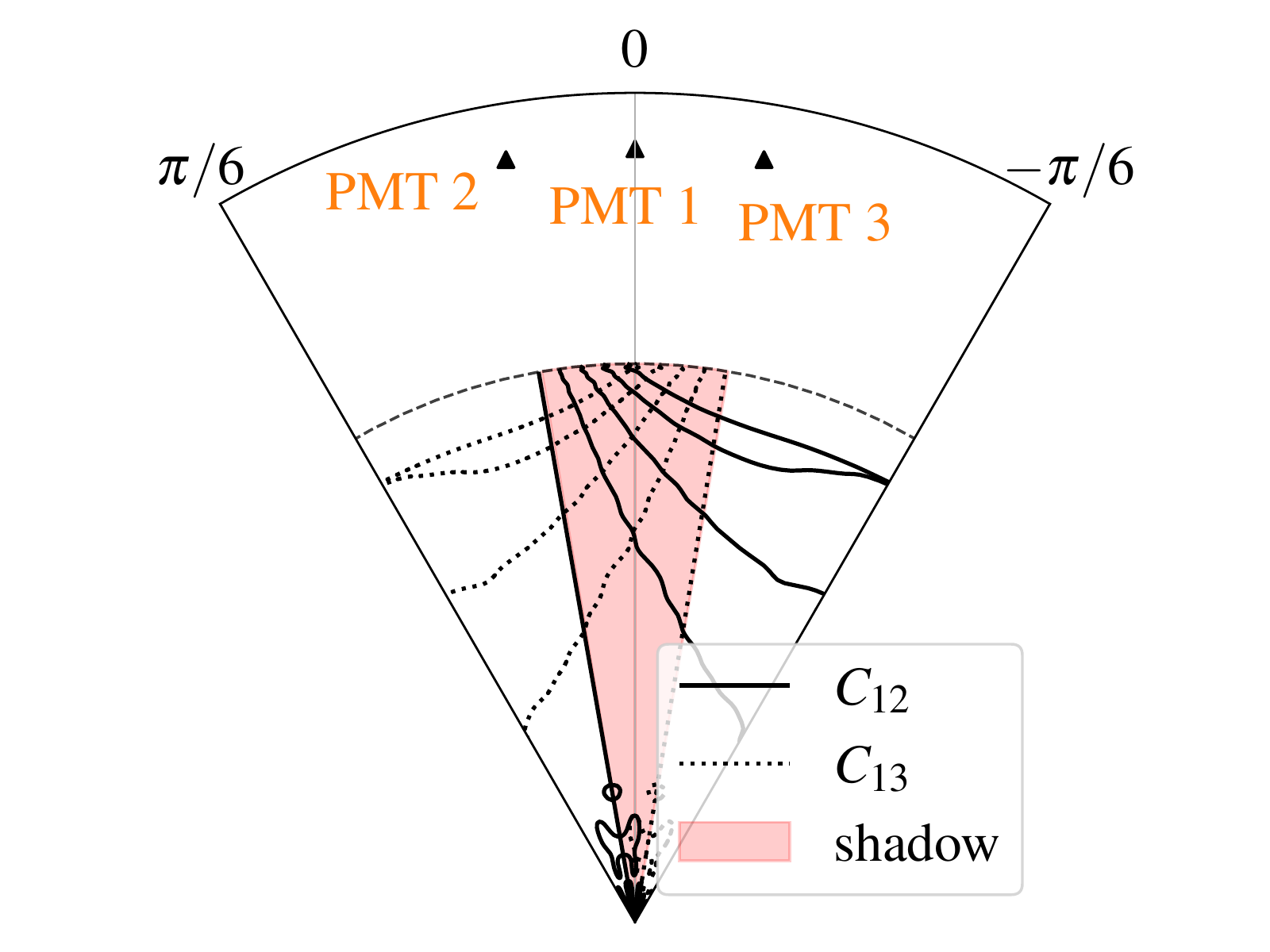}
		\subcaption{Ideal detector arrangement}
		\label{fig: Equiv_ideal}
	\end{minipage}
	\caption{ER lines of different detector response models. PMTs are in black triangles. (\subref{fig: d2}) -- (\subref{fig: ER_100}) are the homogeneous materials with $r_\mathrm{PMT} = r_\mathrm{LS}$. For (\subref{fig: d2}) and (\subref{fig: d2cos}), PMTs are arranged at 0, $2\pi/3$, $4\pi/3$. For (\subref{fig: ER_100}), PMTs are arranged at 0, $\pm \pi/100$. (\subref{fig: Equiv_outlet}) -- (\subref{fig: Equiv_ideal}) are the inhomogeneous materials, using the coefficients shown in figure~\ref{fig: Probe_PE}. (\subref{fig: Equiv_outlet}) is the outlet direction, PMTs are arranged at $2\pi/5$ and $\pm \pi/5$, lacking at 0. The intersections of $C_{12}=1$ and $C_{13}=1$ are at the central and top regions. (\subref{fig: Equiv_general}) is the general direction. PMTs are arranged at $0$ and $\pm \pi/5$. More than one intersections of $C_{12}=1$ and $C_{13}=4.5$ are at \SI{350}{mm} and \SI{550}{mm}. (\subref{fig: Equiv_ideal}) is the ideal detector without multimodality. PMTs are arranged at 0, $\pm \pi/18$. For all subfigures, the vertices are constrained in the shadow region. We circle the multiple intersections of ER lines in (\subref{fig: d2cos}), (\subref{fig: ER_100}), (\subref{fig: Equiv_outlet}) and (\subref{fig: Equiv_general}).}
	\label{fig: ER}
\end{figure}

Observing that closer PMTs contribute more to the likelihood function, we only consider the three closest PMTs to the vertex for simplicity. Considering symmetry, in figure~\ref{fig: ER} the vertices are constrained in the shadow region without loss of generality. We define the \emph{contour function} by
\begin{equation}
	\label{eq: contour}
	C_{1i}(\mathbf{r}) = \frac{\lambda_{1}(\mathbf{r})}{\lambda_{i}(\mathbf{r})},
\end{equation}
where \(\lambda_{1}(\mathbf{r})\) is the expected number of PEs at the closest PMT to $\mathbf{r}$, and \(\lambda_i(\mathbf{r})\) is that of the 2nd or 3rd closest PMT with $i=2$ or $3$ respectively. In a two-dimensional slice of the detector, a certain value of $C_{1i}(\mathbf{r})$ defines an \emph{equivalent ratio} (ER) line. If one ER line of $C_{12}$ and another in $C_{13}$ has two intersections $\mathbf{r}_1$ and $\mathbf{r}_2$, $d_\mathrm{cos}\left[\Lambda(\mathbf{r}_1) \mid \Lambda(\mathbf{r}_2)\right] \approx 0$. Because the 3 dominanting PMTs satisfies $\lambda_1(\mathbf{r}_1):\lambda_2(\mathbf{r}_1):\lambda_3(\mathbf{r}_1) = \lambda_1(\mathbf{r}_2):\lambda_2(\mathbf{r}_2):\lambda_3(\mathbf{r}_2)$.  Therefore it is the key to count the intersections between any pair of $C_{12}$ and $C_{13}$ ER lines.

We study the multimodality affected by angle of incidence $\beta$ for \emph{homogeneous materials} and TR for \emph{inhomogeneous materials} respectively.  In the former case, the LS and the buffer have the same refractive index. Assuming the PMTs are small enough and their surface is flat, the predicted PE $\lambda_i$ is proportional to the solid angle $\Omega_i$. $\Omega_i$ is subtended by the $i$th PMT from the vertex,
\begin{equation}
	\label{eq: d2cos}
	\lambda_i(\mathbf{r}) \propto \Omega_i(\mathbf{r}) \propto \frac{1}{{\lVert \mathbf{r} - \mathbf{r}_{\mathrm{PMT}, i} \rVert}_2^2}\cos\beta_i,
\end{equation}
where $\mathbf{r}$, $\mathbf{r}_{\mathrm{PMT}, i}$ are defined in figure~\ref{fig: SH}. $\beta$ is the angle of incidence on a PMT. Eq.~\eqref{eq: d2cos} well describes TAO~\cite{abusleme2020tao} to be equipped with SiPM and XMASS~\cite{abe2013xmass} with flat-photocathode R10789 PMTs~\cite{abe_development_2019}. To comparatively study the effect by $\beta_i$ by leaving it out,
\begin{equation}
	\label{eq: d2}
	\lambda_i(\mathbf{r}) \propto \Omega_i(\mathbf{r}) \propto \frac{1}{{\lVert \mathbf{r} - \mathbf{r}_{\mathrm{PMT}, i} \rVert}_2^2}.
\end{equation}
We set $r_\mathrm{PMT} = r_\mathrm{LS}$, namely no buffer is installed between LS and PMTs.  The PMTs located at $0$, $2\pi/3$ and $4\pi/3$ directions are indexed by 1, 2 and 3. $C_{12}$ and $C_{13}$ are symmetric. Figure~\ref{fig: d2cos} demonstrates the ER lines with \(\beta\) and figure~\ref{fig: d2} demonstrates those without.  The two intersections circled red in figure~\ref{fig: d2cos} show that $\beta$ leads to multimodality.

To account for the suspicion that multimodality is induced by sparse PMT arrangement in figure~\ref{fig: d2cos}, consider two intersections with a more compact arrangement under eq. \eqref{eq: d2cos} in figure~\ref{fig: ER_100}.  Regardless of the PMT arrangement, the shadow boundary is not only an ER line but also the bisector of the two neighboring PMTs. The point on the bisector at $r = r_\mathrm{LS}$ is always one of the intersections under eq. \eqref{eq: d2cos}. PMT is almost blind to this point due to a large $\beta$. To avoid this, a non-fluorescent buffer is necessary to isolate scintillation events away from the PMTs.

We use \emph{inhomogeneous materials} to study the TR effect. The buffer's refractive index is different from the central detector. For the Jinping prototype, we see from the fitted model that TR makes ER lines segmented.  A two-dimensional slice of the Jinping prototype includes 10 PMTs. We pick the 3 PMTs at $(0, \pm{\pi/5})$ to represent the \emph{general direction} in figure~\ref{fig: Equiv_general}, matching figure~\ref{fig: plane_zoom}. It is evident from figure~\ref{fig: Equiv_general} that events at \SI{350}{mm} and \SI{550}{mm} degenerate. We also pick $(\pm{\pi/5}, 2\pi/5)$ as the \emph{outlet direction} in figure~\ref{fig: Equiv_outlet}, matching figure~\ref{fig: cosdist_z}. Vertices at $z=$ \SI{600}{mm} and the center degenerate.  Such 3-PMT plots are reasonable simplifications to reproduce the trends of $d_{\cos}$ in all-PMT cases.

The ER lines near the TR region are distorted compared to the homogeneous materials. The rich functional structures in this region are a hotbed to multimodality. A solution is to reduce the gaps between PMTs, making them close enough so as not to fall into each other's TR regions. For example, the ideal detector has no multiple intersections in ER lines in figure~\ref{fig: Equiv_ideal}.

\subsection{A criterion against multimodality}
\label{subsec: criterion}

In figure~\ref{fig: d2}, the predicted PE is approximately proportional to the inverse square of the distance from the vertex to PMT. The incident angle and the TR break that trend. These effects are so strong that the closest PMTs could receive fewer PEs, making the ER lines distorted and intersect multiple times. They are the seeds to multimodality.  Therefore a perfect inverse-squared detector in figure~\ref{fig: d2} is always free from multimodality.

Taking $\lambda_1$ and $\lambda_2$ to be the same meaning as eq.~\eqref{eq: contour}, in figures~\ref{fig: d2cos} and \ref{fig: ER_100}, adding buffer is equivalent to reducing $\lambda_1$.  Conditions of figures~\ref{fig: Equiv_outlet} and \ref{fig: Equiv_general} improve by reducing the gap between neighboring PMTs, making $\lambda_2$ larger. The two phenomena can be unified by $\lambda_1/\lambda_2$: for any event, $\lambda_1/\lambda_2<10$ is required to avoid multimodality.

\begin{figure}[!htbp]
	\centering
	\begin{minipage}[htb]{0.45\textwidth}
		\centering
		\includegraphics[trim=0 10 10 0, clip, width=\linewidth]{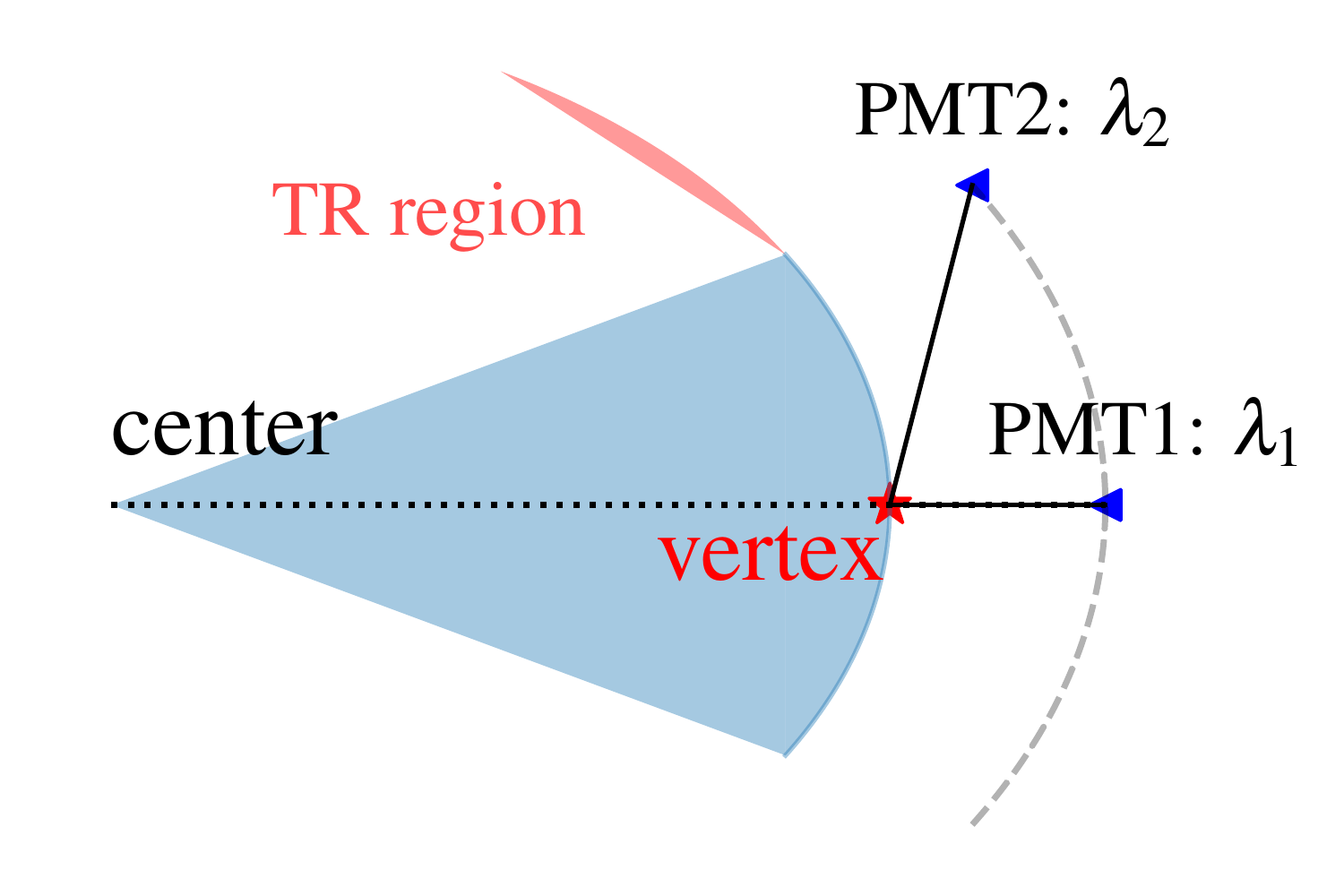}
		\subcaption{The criterion $\lambda_1/\lambda_2>10$.}
		\label{fig: criterion}
	\end{minipage}
	\quad
	\begin{minipage}[htb]{0.45\textwidth}
		\centering
		\includegraphics[trim=0 10 10 0, clip, width=\linewidth, page=10]{equiv.pdf}
		\subcaption{Least needed number of the PMTs based on the refractive index of the outer buffer and $r_\mathrm{PMT}$}
		\label{fig: PMT_coverage}
	\end{minipage}

	\caption{(\subref{fig: criterion}) shows the criterion to avoid multimodality. $\lambda_1/\lambda_2$ should be less than 10 and the TR region should be excluded. (\subref{fig: PMT_coverage}) shows the least number of needed PMTs. $r_\mathrm{LS}$ is fixed to \SI{645}{mm}. For the Jinping prototype, the average refractive index of the water is 1.33 and $r_\mathrm{PMT}$ is \SI{832}{mm}, showing that at least 120 PMTs are necessary. The red star is the ideal detector with $r_\mathrm{PMT}=$~\SI{900}{mm}, requiring at least 90 PMTs. The larger the $r_\mathrm{PMT}$, the refractive index and the number of PMTs help to avoid multimodality. }
\end{figure}

It is sufficient to examine if an extreme vertex with the biggest $\lambda_1/\lambda_2$ is less than 10.  Such an extreme vertex is approximately realized in the center-PMT direction at $r=r_\mathrm{LS}$, illustrated in figure~\ref{fig: criterion}.  Note that $\lambda_1/\lambda_2<10$ is a necessary condition, we might construct some case where the TR region is fully contained in the gap between 2 PMTs to embed a lot of degeneracy but still satisfies $\lambda_1/\lambda_2<10$. $\lambda_1$ and $\lambda_2$ is related to the $r_\mathrm{LS}$, $r_\mathrm{PMT}$, $N_\mathrm{PMT}$ and the buffer's refractive index.

Keeping an identical distance between neighboring PMTs to extend our criterion to three dimensions, $N_\mathrm{PMT, 3D}$ scales approximately as $N_\mathrm{PMT, 2D}^2/3$.  Fixing $r_\mathrm{LS} = \SI{650}{mm}$, figure~\ref{fig: PMT_coverage} shows the least needed number of PMTs under different $r_\mathrm{PMT}$ and the buffer refractive index. For a certain $N_\mathrm{PMT}$, larger $r_\mathrm{PMT}$ or the refractive index gives better resilience against degeneracy. For $r_\mathrm{PMT}=$~\SI{832}{mm} and the refractive index 1.33, 120 PMTs are needed. Using 8-inch PMTs, the PMT coverage, calculated by $N_\mathrm{PMT} \times (10\mathrm{mm})^2 / 4 r^2_\mathrm{PMT}$, should exceed \SI{40}{\%}. Be aware that the larger $r_\mathrm{PMT}$ leads to poorer PMT coverage, thus $r_\mathrm{PMT}$ should be as close to the lower limit in figure~\ref{fig: PMT_coverage} as possible. $r_\mathrm{PMT}$'s other influences will be discussed in section~\ref{subsec: Other reflections by acrylic shell}.

The ideal detector with $r_\mathrm{PMT}=$~\SI{832}{mm} requires at least 90 PMTs. To reduce the influences of fluctuations, we use 120 PMTs to guarantee perfect reconstruction performance.

\section{Reconstruction results}
\label{sec4}
We verify our vertex reconstruction on MC and raw data. For MC, we use the Jinping prototype to verify the conditions of multimodality in section~\ref{subsec: criterion}, and the ideal detector to study the reconstruction bias and resolution. The simulations use \SI{2}{MeV} $e^{-}$ on $x$ and $z$ axes: $x$ represents the general direction, and $z$ the outlet direction, aligned with the definitions in section~\ref{subsec: Cosine distance for 3-PMT case}. For raw data, we analyze the \ce{^{214}Bi}-\ce{^{214}Po} cascade signals by the model fitted from the Jinping-prototype simulation.

\subsection{Results from simulations}
\label{subsec: MC data results}
We compare the results using MLE and BC. For BC, the reconstructed vertex $\hat{\mathbf{r}}$ is
\begin{equation}
	\hat{\mathbf{r}} = 1.5 \times \frac{\sum_i q_i \mathbf{r}_{\mathrm{PMT}, i}}{\sum_i q_i}
	\label{eq: BC}
\end{equation}
and the energy is scaled from the total number of hits,
\begin{equation}
	\frac{\hat{E}}{[\mathrm{MeV}]} = \frac{N_\mathrm{hit}}{65}
	\label{eq: scaled_total_PE}
\end{equation}
where $1.5$ is a correction factor and 65 is the average total PE at \SI{1}{MeV} at the Jinping prototype. The simulation data ranges from \SIrange{0}{650}{mm} with steps of \SI{10}{mm}. Each step contains 5000 events.
\begin{figure}[!htbp]
	\centering
	\begin{minipage}[htb]{0.242\linewidth}
		\centering
		\includegraphics[width=1\linewidth, page=3, trim=0 15 0 0, clip]{Recon_h.pdf}
		\subcaption{MLE}
		\label{fig: SH_0.83_x}
	\end{minipage}
	\begin{minipage}[htb]{0.242\linewidth}
		\centering
		\includegraphics[width=1\linewidth, page=7, trim=0 15 0 0, clip]{Recon_h.pdf}
		\subcaption{BC}
		\label{fig: BC_0.83_x}
	\end{minipage}
	\begin{minipage}[htb]{0.242\linewidth}
		\centering
		\includegraphics[width=1\linewidth, page=4, trim=0 15 0 0, clip]{Recon_h.pdf}
		\subcaption{MLE}
		\label{fig: SH_0.83_z}
	\end{minipage}
	\begin{minipage}[htb]{0.242\linewidth}
		\centering
		\includegraphics[width=1\linewidth, page=8, trim=0 15 0 0, clip]{Recon_h.pdf}
		\subcaption{BC}
		\label{fig: BC_0.83_z}
	\end{minipage}
	\\
	\begin{minipage}[htb]{0.48\textwidth}
		\centering
		\includegraphics[width=1\linewidth, page=3, trim=0 15 0 0, clip]{Energy.pdf}
		\subcaption{Energy reconstruction on $x$ axis}
		\label{fig: energy_0.83_x}
	\end{minipage}
	\quad
	\begin{minipage}[htb]{0.48\textwidth}
		\centering
		\includegraphics[width=1\linewidth, page=4, trim=0 15 0 0, clip]{Energy.pdf}
		\subcaption{Energy reconstruction on $z$ axis}
		\label{fig: energy_0.83_z}
	\end{minipage}
	\hfill
	\begin{minipage}[htb]{0.242\textwidth}
		\centering
		\includegraphics[width=1\linewidth, page=1, trim=0 15 0 0, clip]{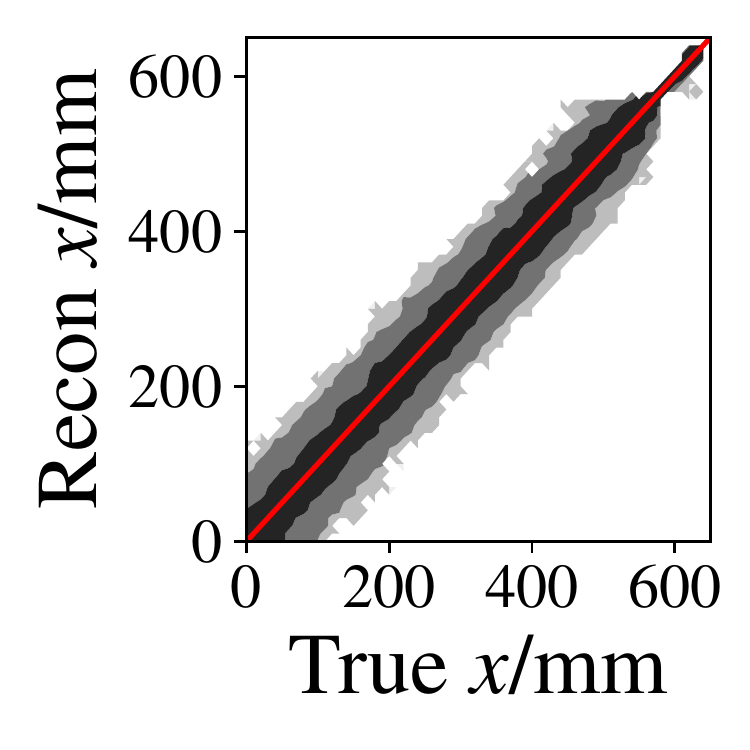}
		\subcaption{MLE}
		\label{fig: SH_0.90_x}
	\end{minipage}
	\begin{minipage}[htb]{0.242\textwidth}
		\centering
		\includegraphics[width=1\linewidth, page=5, trim=0 15 0 0, clip]{Recon_h.pdf}
		\subcaption{BC}
		\label{fig: BC_0.90_x}
	\end{minipage}
	\begin{minipage}[htb]{0.242\textwidth}
		\centering
		\includegraphics[width=1\linewidth, page=2, trim=0 15 0 0, clip]{Recon_h.pdf}
		\subcaption{MLE}
		\label{fig: SH_0.90_z}
	\end{minipage}
	\begin{minipage}[htb]{0.242\textwidth}
		\centering
		\includegraphics[width=1\linewidth, page=6, trim=0 15 0 0, clip]{Recon_h.pdf}
		\subcaption{BC}
		\label{fig: BC_0.90_z}
	\end{minipage}
	\\
	\begin{minipage}[htb]{0.48\textwidth}
		\centering
		\includegraphics[width=1\linewidth, page=1, trim=0 15 0 0, clip]{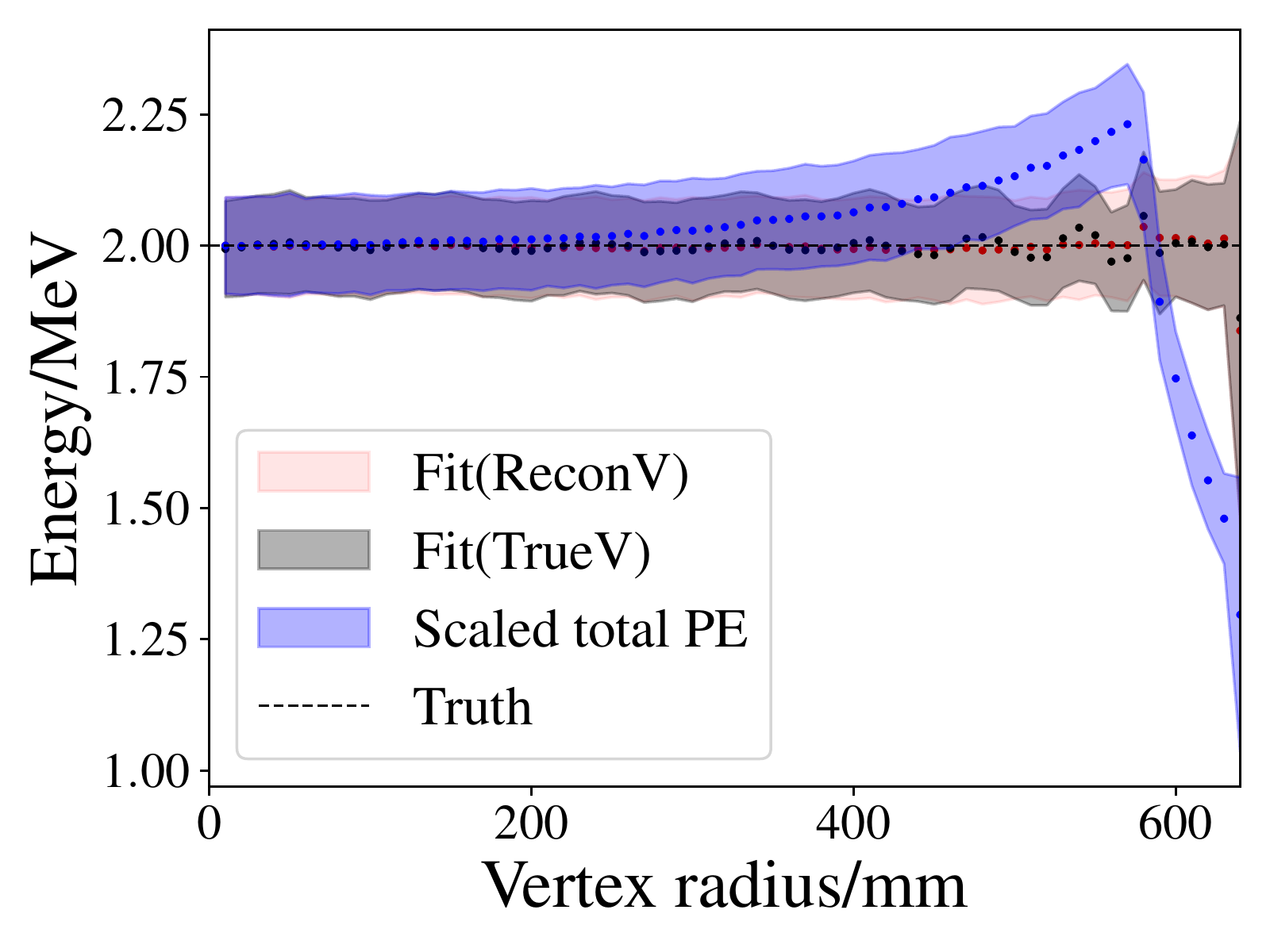}
		\subcaption{Energy reconstruction on $x$ axis}
		\label{fig: energy_0.90_x}
	\end{minipage}
	\quad
	\begin{minipage}[htb]{0.48\textwidth}
		\centering
		\includegraphics[width=1\linewidth, page=2, trim=0 15 0 0, clip]{Energy.pdf}
		\subcaption{Energy reconstruction on $z$ axis}
		\label{fig: energy_0.90_z}
	\end{minipage}
	\caption{Vertex and energy reconstruction performance of \SI{2}{MeV} $e^{-}$. (\subref{fig: SH_0.83_x}) -- (\subref{fig: energy_0.83_z}) are from the Jinping prototype.  The red dashed lines in (\subref{fig: SH_0.83_x}) and (\subref{fig: SH_0.83_z}) show the degeneracies. (\subref{fig: SH_0.90_x}) -- (\subref{fig: energy_0.90_z}) are from the ideal detector. In (\subref{fig: SH_0.83_x}) -- (\subref{fig: BC_0.83_z}) and (\subref{fig: SH_0.90_x}) -- (\subref{fig: BC_0.90_z}), the light grey, grey and black represent bins with 1--10, 10--100 and $>100$ events, respectively. (\subref{fig: energy_0.83_x}), (\subref{fig: energy_0.83_z}) and (\subref{fig: energy_0.90_x}), (\subref{fig: energy_0.90_z}) are the energy reconstruction results of \SI{2}{MeV} $e^{-}$ on the Jinping prototype and the ideal detector. ``Fit(ReconV)'' and ``Fit(TrueV)'' are energies with reconstructed and true vertices, respectively. ``Scaled total PE'' is by eq. \eqref{eq: scaled_total_PE}.}
	\label{fig: Recon}
\end{figure}

Figure~\ref{fig: SH_0.83_x} shows a reasonable vertex reconstruction in the general direction.  That is in contrast to the outlet direction in figure~\ref{fig: SH_0.83_z}, where events at the detector center and around $(0, 0, 600)$~\si{mm} are indistinguishable. Similarly, events at \SI{350}{mm} and \SI{550}{mm} degenerate.  Such degeneracy is a consequence of multimodality in the likelihood function as discussed in section~\ref{subsec: criterion}.  Notice that the $d_{\cos}$ relations in figures~\ref{fig: cosdist_x} and \ref{fig: cosdist_z} predict the major patterns of MLE reconstructions in figures~\ref{fig: SH_0.83_x} and \ref{fig: SH_0.83_z}.

In figures~\ref{fig: BC_0.83_x} and \ref{fig: BC_0.83_z}, similar worsen trends appear for BC from the general to the outlet direction. They both have severe biases.

For the ideal detector, no outlet is considered.  $x$ and $z$ axes are equivalent.  MLE performs well without degeneracy in figures~\ref{fig: SH_0.90_x} and \ref{fig: SH_0.90_z}, proving the effectiveness of the $\lambda_1/\lambda_2 < 10$ criteria in section~\ref{subsec: criterion}.  BC, on the contrary, still reconstructs badly in the TR region.  Our detector model in section~\ref{sec2} describes the TR region well and consequently MLE provides a big improvement upon the BC method.

The energy reconstruction for the Jinping detector is shown in figure~\ref{fig: energy_0.83_x} and \ref{fig: energy_0.83_z}. The scaled total PE decreases rapidly in the TR region, which introduces a big bias. The reconstructed energy by MLE is even worse, due to the wrong vertices. The true vertices give an unbiased energy estimation, indicating that the detector response model is accurate. For the ideal detector of figure~\ref{fig: energy_0.90_x} and \ref{fig: energy_0.90_z}, BC has similar trends. The MLE is almost unbiased and improved by eliminating the vertex degeneracy.

Excluding the regions $| \hat{\mathbf{r}} | > $ \SI{600}{mm} and $\sqrt{\hat{x}^2 + \hat{y}^2}<$ \SI{50}{mm}, the standard deviation of Jinping prototype is \SI{60}{mm} at the center and \SI{20}{mm} in the TR region. The energy reconstruction heavily depends on the vertex. When using the true vertex, the results are almost unbiased even in the TR region. The energy resolution of the Jinping prototype is approximately \SI{11}{\%}$/\sqrt{E\mathrm{[MeV]}}$.

\subsection{Results from raw data}
\label{subsec: Raw data results}
The raw data of the Jinping prototype is the PMT waveform $w_i(t)$, which is a convolution of hits $\sum_j \delta(t-t_{ij})$ and the \emph{single PE response} $V_\mathrm{PE}(t)$,
\begin{equation}
	w_i(t) = \sum_j\delta(t-t_{ij})\otimes V_\mathrm{PE}(t)+\epsilon(t) = \sum_j V_\mathrm{PE}(t - t_{ij})+\epsilon(t),
\end{equation}
$\epsilon(t)$ is Gaussian white noise. We use \emph{Richardson-Lucy direct demodulation} (LucyDDM) for deconvolution~\cite{richardson1972bayesian}. The input is the timing series $t_{ij}$, returning the gain modified charge $q_{ij}$. Traditional LucyDDM is biased in photon density due to an artificial threshold. Xu et al.~\cite{xu2021towards} use a rescaling factor to significantly reduce the bias. In terms of the photon density resolution, LucyDDM matches the fit method~\cite{xu2021towards} and consumes less time.

Xu et al.~\cite{xu2021towards} also utilize the non-normalized Kullback-Leibler (KL) divergence~\cite{Kullback1951, 10.2307/2337385} for reconstruction, which is a special case of \emph{density power divergence}~\cite{10.2307/2337385}. It joins the waveform analysis results $\sum_j q_{ij}\delta(t - t_{ij})$ and the inhomogeneous Poisson process. We take the non-normalized KL divergence as the pseudo-minus-log-likelihood in this work.
\begin{equation}
	\begin{aligned}
		- \log\mathscr{L}(E, \mathbf{r}, t_0) & = \sum_i D_\mathrm{KL}\left[\sum_j q_{ij}\delta(t - t_{ij}) \;\middle|\; E \lambda_i(r,\theta_i) R(t - t_0 - T_i(r,\theta_i))\right]                                                                               \\
		                                      & = - \sum_i\left[\sum_j q_{ij} \log R(t_{ij} - t_0 - T_i) + \left(\sum_j q_{ij}\right)\log (E\lambda_i) - E\lambda_i \right]+ \mathrm{Const.}                                                                       \\
		                                      & \sim \overbrace{\sum_i\sum_j q_{ij} \frac{-\mathscr{R}_\tau(t_{ij} - T_i - t_0)}{t_\mathrm{s}}}^{\mathrm{timing\ part}} + \overbrace{\sum_i \left[-q_i\log (E\lambda_i) + E\lambda_i \right]}^{\mathrm{PE\ part}},
		\label{eq: Wave_Likelihood}
	\end{aligned}
\end{equation}
where $q_i = \sum_j q_{ij}$.

\begin{figure}[H]
	\centering
	\begin{minipage}[htb]{0.45\textwidth}
		\centering
		\includegraphics[width=1\linewidth, trim=10 10 10 0, clip, page=1]{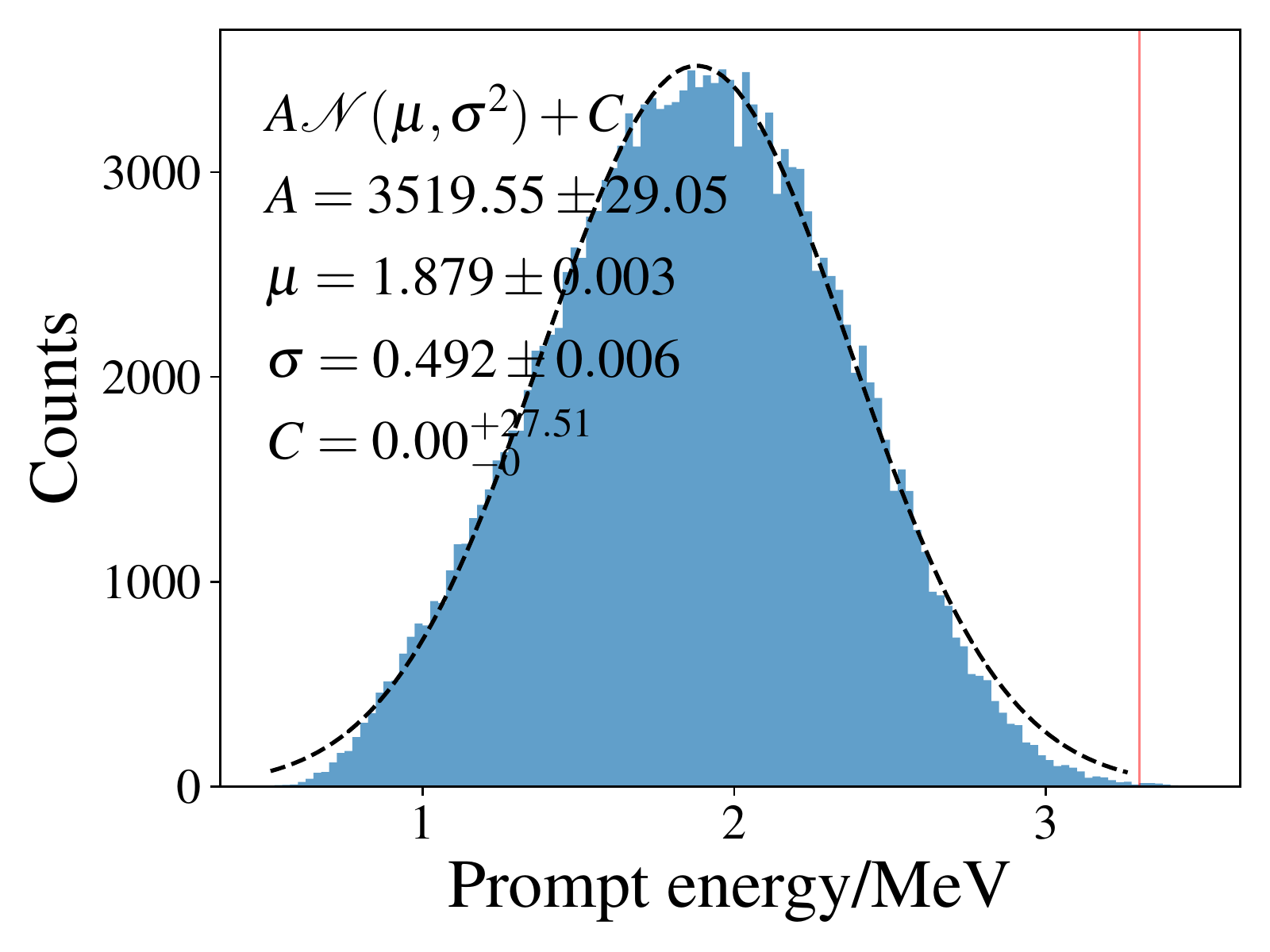}
		\subcaption{Prompt energy cut}
		\label{fig: raw_a}
	\end{minipage}
	\quad
	\begin{minipage}[htb]{0.45\textwidth}
		\centering
		\includegraphics[width=1\linewidth, trim=10 10 10 0, clip, page=2]{ReconReal.pdf}
		\subcaption{Delayed energy cut}
		\label{fig: raw_b}
	\end{minipage}
	\hfill
	\begin{minipage}[htb]{0.45\textwidth}
		\centering
		\includegraphics[width=1\linewidth, trim=10 10 10 0, clip, page=3]{ReconReal.pdf}
		\subcaption{Delayed time cut}
		\label{fig: raw_c}
	\end{minipage}
	\quad
	\begin{minipage}[htb]{0.45\textwidth}
		\centering
		\includegraphics[width=1\linewidth, trim=10 10 10 0, clip, page=4]{ReconReal.pdf}
		\subcaption{Distance cut}
		\label{fig: raw_d}
	\end{minipage}
	\caption{Reconstructions of \ce{^{214}Bi}-\ce{^{214}Po} cascades in raw data. $\mathscr{N}(\mu, \sigma^2)$ is a Gaussian distribution with the mean $\mu$ and variance $\sigma^2$. (\subref{fig: raw_a}) and (\subref{fig: raw_b}) is fitted by Gaussian. (\subref{fig: raw_c}) is fitted by an exponential distribution.}
	\label{fig: 4-1 plot}
\end{figure}

The detector response model is based on simulation since there is no dedicated calibration runs~\cite{zhao2022measurement}. We check the \ce{^{214}Bi}-\ce{^{214}Po} cascade signal of $e^-$ prompt and $\alpha$ delayed. The cuts are listed below:
\begin{enumerate}
	\item $|\hat{\mathbf{r}}_\mathrm{prompt}|<\SI{600}{mm}$ and $|\hat{\mathbf{r}}_\mathrm{delay}| < \SI{600}{mm}$,
	\item $\sqrt{\hat{x}_\mathrm{prompt}^2 + \hat{y}_\mathrm{prompt}^2} > \SI{50}{mm}$ and $\sqrt{\hat{x}_\mathrm{delay}^2 + \hat{y}_\mathrm{delay}^2} > \SI{50}{mm}$,
	\item Visible energy of $e^{-}$ is less than \SI{3.5}{MeV},
	\item Visible energy of $\alpha$ is in \SIrange{0.7}{1}{MeV},
	\item Delayed time between the prompt and delayed signal in \SIrange{10}{1000}{us},
	\item $| \hat{\mathbf{r}}_\mathrm{prompt} - \hat{\mathbf{r}}_\mathrm{delay} | < \SI{300}{mm}$.
\end{enumerate}
Figures~\ref{fig: raw_a},~\ref{fig: raw_b},~\ref{fig: raw_c} and \ref{fig: raw_d} show data distribution of the six cuts except the 3rd, 4th, 5th and 6th, respectively. The cut is colored red in each subfigure.

Most backgrounds are gammas from the PMTs. For simplicity, we treat it as a constant. The 25-PMT threshold distorts the beta decay energy spectrum. We use Gaussian to fit the prompt and delayed signals. The average energy of the $e^{-}$ is approximately $\sim$ \SI{1.879}{MeV}, and the $\alpha$ peaks at \SI{0.765}{MeV} due to the ionization quenching. The fitted half-life is close to the true value (\SI{164.3}{us}). The results show that most selected events are cascade signals and our reconstruction works well.

\section{Discussion}
\label{sec5}

\subsection{Other reflections by acrylic shell}
\label{subsec: Other reflections by acrylic shell}
The closer PMT often has a larger gradient in the likelihood function. We select the three closest PMTs in section~\ref{subsec: Cosine distance for 3-PMT case} since they are more sensitive to vertex positions. In inhomogeneous materials, normal reflections other than TR at the media boundary focus light, making predicted PEs peak at some regions that are otherwise as dim as their neighbors. Such focal points and lines are sources to multimodality in the likelihood function.

\begin{figure}[!htbp]
	\centering
	\begin{minipage}[htb]{0.36\textwidth}
		\centering
		\includegraphics[width=\linewidth]{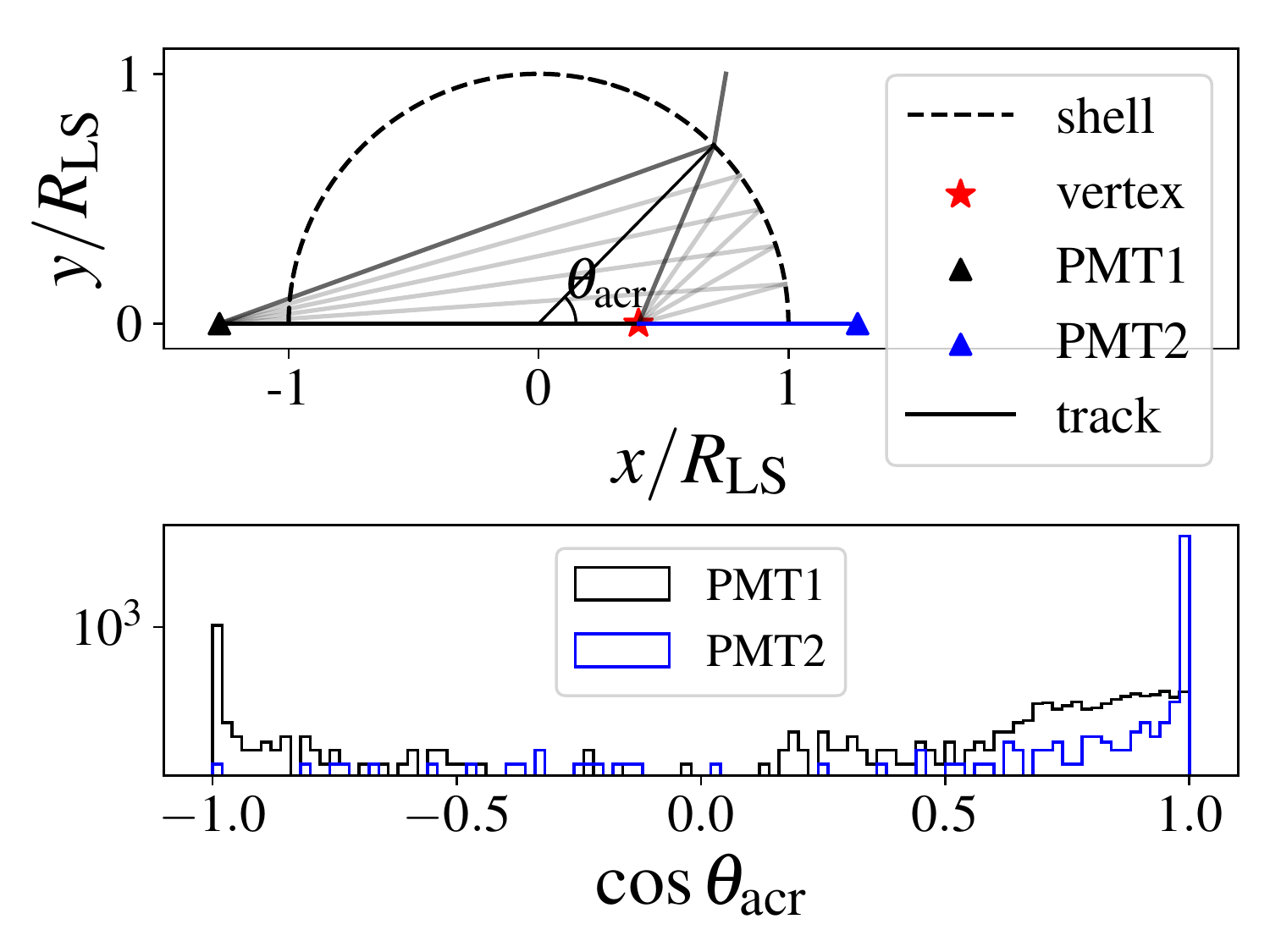}
		\subcaption{Photon tracks at $r=$ \SI{260}{mm}}
		\label{fig: focus}
	\end{minipage}
	\begin{minipage}[htb]{0.31\textwidth}
		\centering
		\vspace{1.5em}
		\includegraphics[width=\linewidth]{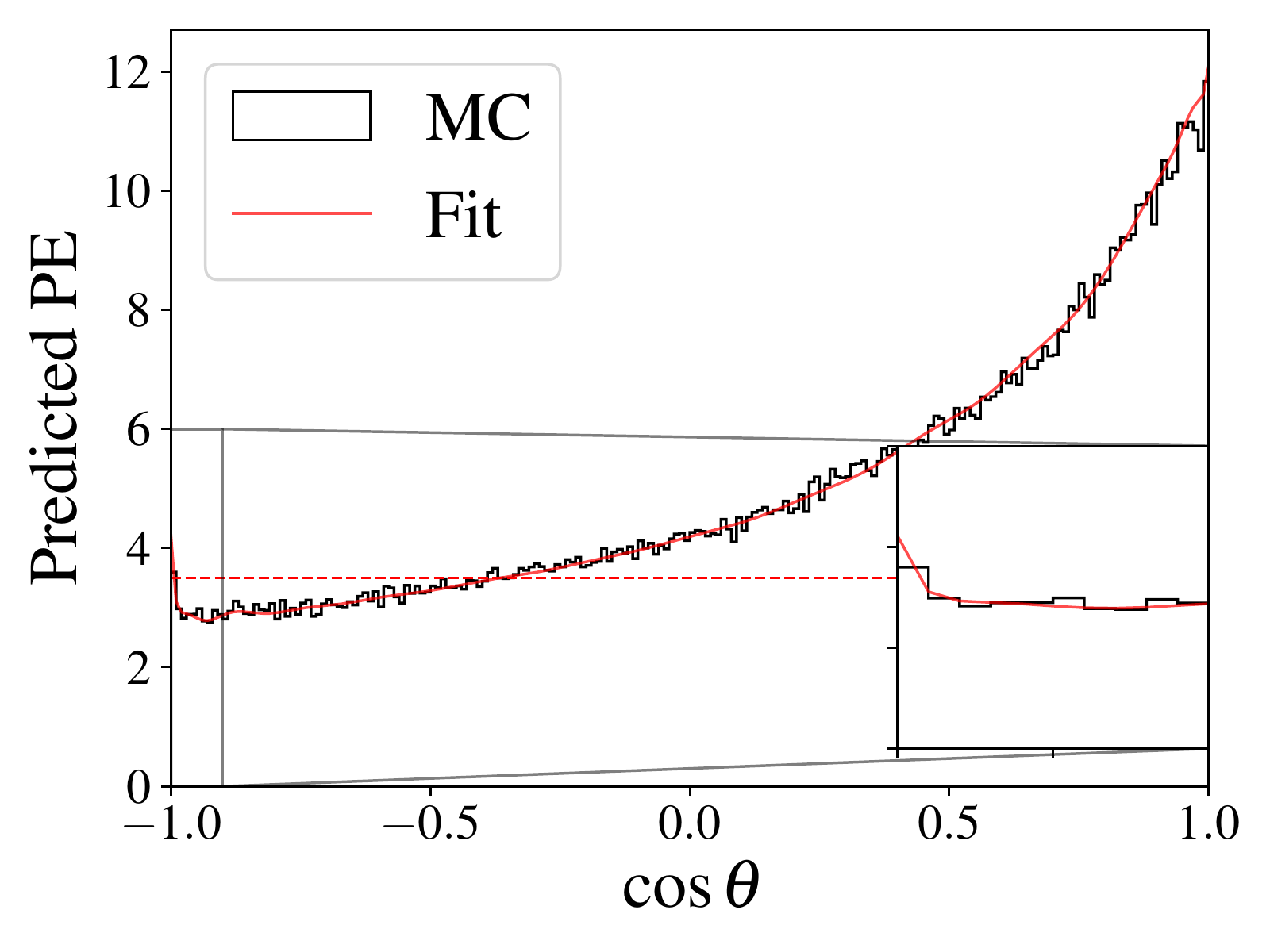}
		\subcaption{Predicted PE at $r=$ \SI{260}{mm}}
		\label{fig: predict}
	\end{minipage}
	\begin{minipage}[htb]{0.31\textwidth}
		\centering
		\vspace{1.5em}
		\includegraphics[width=\linewidth, page=1]{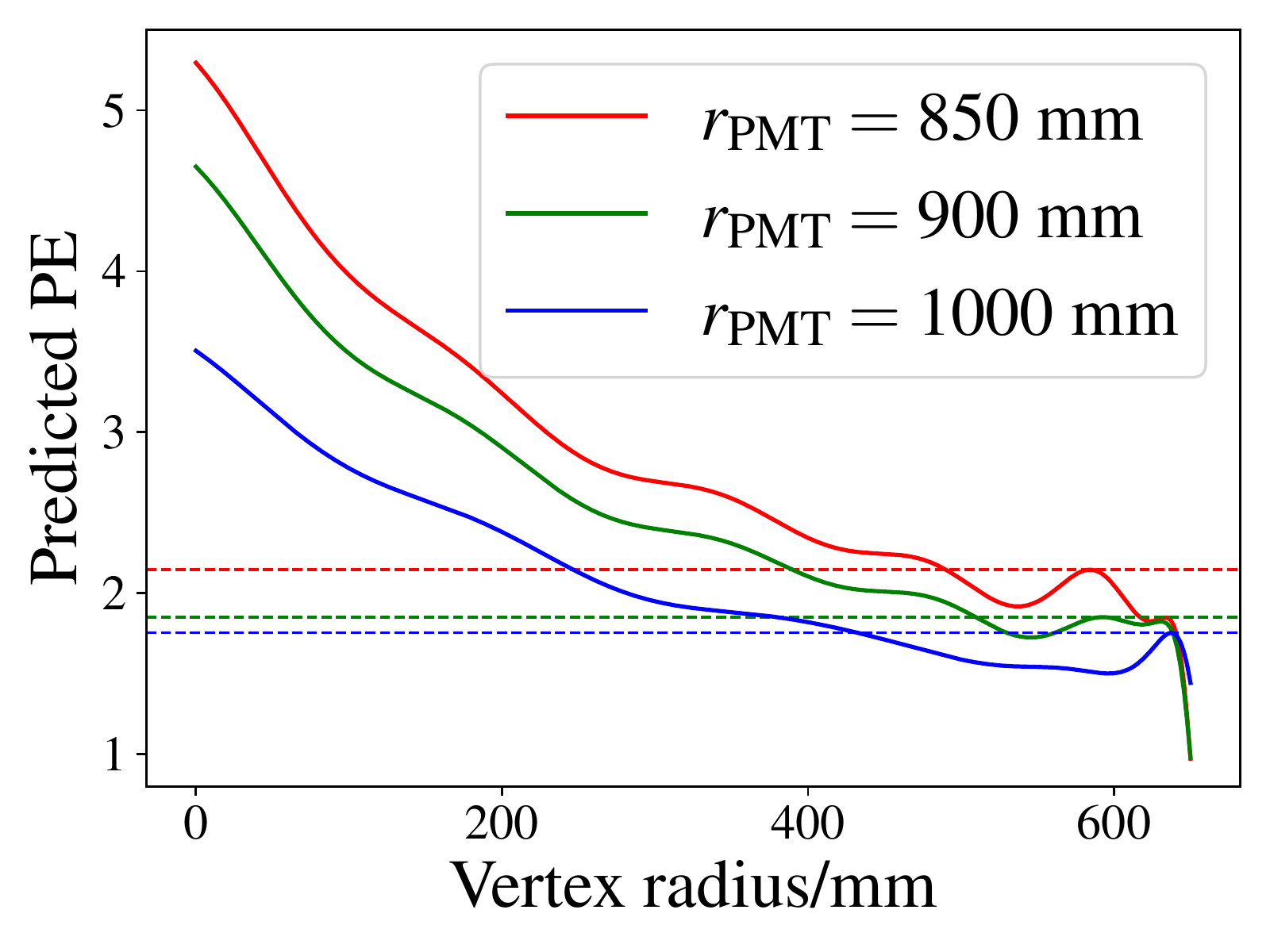}
		\subcaption{Predicted PE at $\theta=\frac{5}{6}\pi$}
		\label{fig: 5/6}
	\end{minipage}
	\caption{Focus effect at $r = 0.4r_\mathrm{LS} =$~\SI{260}{mm}. The tracks of detected photons are sketched in (\subref{fig: focus}).  The top of (\subref{fig: focus}) defines $\theta_\mathrm{acr}$ of the tracks.  The vertex is relative to PMT1 at $\cos\theta_1=-1$ and PMT2 at $\cos\theta_2=1$.  The bottom of (\subref{fig: focus}) is the log-scale distribution of $\cos\theta_\mathrm{acr}$. (\subref{fig: predict}) is a slice of $r=0.4r_\mathrm{LS}$ shown in figure~\ref{fig: Probe_PE}. ``MC'' is the histogram of the simulated events, and ``Fit'' represents the fitted response in section~\ref{sec2}. The inlet zooms in the peak at $\cos\theta = -1$. (\subref{fig: 5/6}) is the predicted PE at $\theta=5\pi/6$ conditioned to different $r_\mathrm{PMT}$. Each dashed line in (\subref{fig: predict}) and (\subref{fig: 5/6}) covers a degenerate region.}
\end{figure}

When the vertices $r/r_\mathrm{LS}\in[0.4,0.5]$, figure~\ref{fig: focus}~(top) shows the photons reflected by the acrylic shell are focused at $\cos\theta \simeq -1$. We record $\theta_\mathrm{acr}$, which is the central angle defined by the vertex and the point of incidence to the acrylic shell, and compare the $\theta_\mathrm{acr}$'s distribution of blue PMT at $\theta \simeq 0$ and black PMT at $\theta \simeq \pi$. In figure~\ref{fig: focus}~(bottom), most photons hit the blue PMT directly, but quite a lot of photons bounce to the black PMT after reflection by the acrylic shell. $r=0.4r_\mathrm{LS}$ is therefore denoted as the \emph{focus region}. In figure~\ref{fig: predict}, the slice of the predicted PE at $r=0.4r_\mathrm{LS}$ peaks at $\cos\theta \simeq -1$, introducing degeneracy of PE predictions. It is also observed at the Borexino CTF~\cite{alimonti2000light}. The dashed horizontal line drawn at the abnormal $\cos\theta \simeq -1$ peak intersects with the PE prediction line, which defines a \emph{degenerate region}.

\begin{figure}[!htbp]
	\centering
	\begin{minipage}[htb]{0.45\textwidth}
		\centering
		\includegraphics[width=\linewidth, trim=10 10 0 0, clip, page=1]{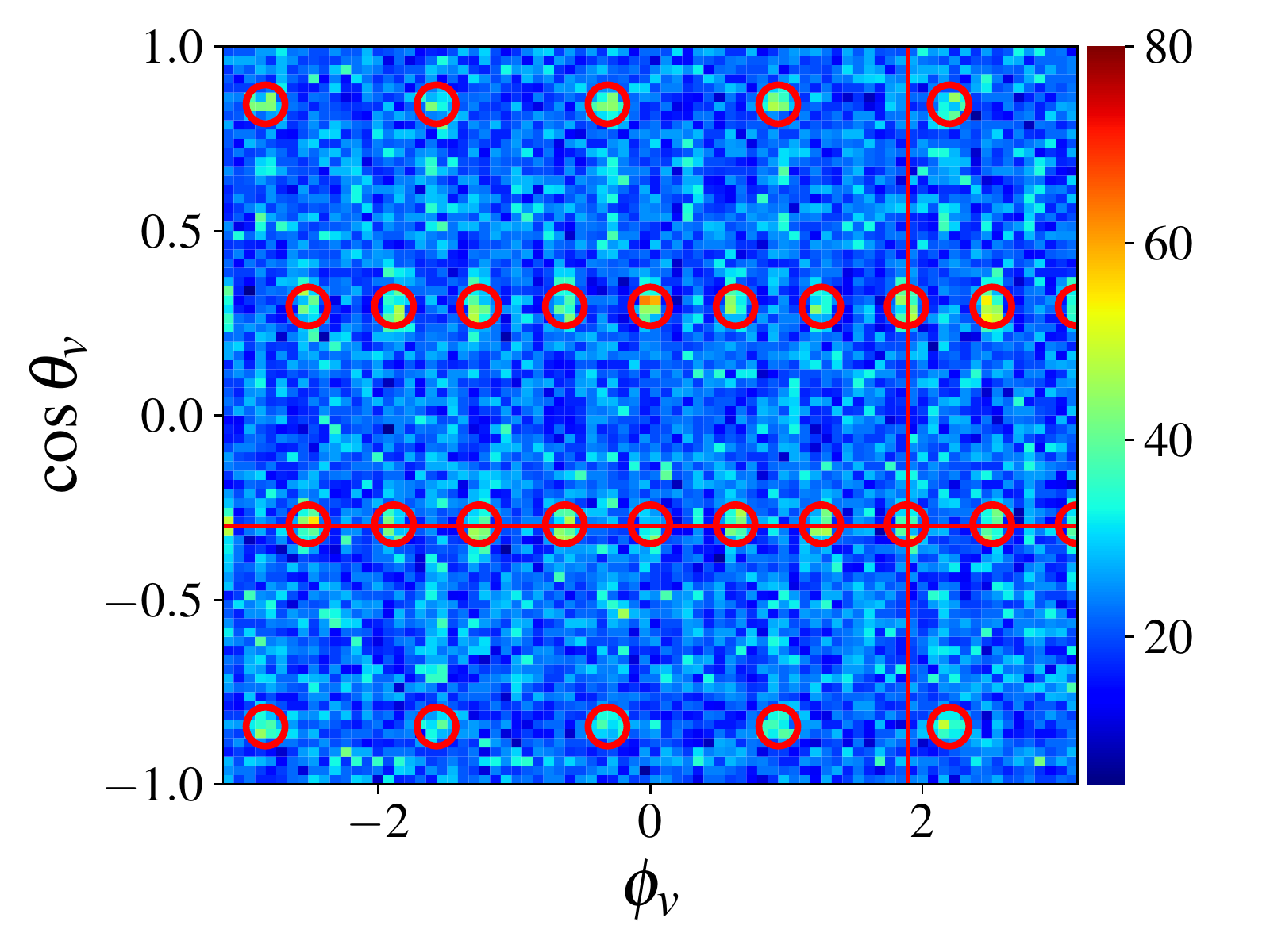}
		\subcaption{Recon with all PMTs}
		\label{fig: all}
	\end{minipage}
	\quad
	\begin{minipage}[htb]{0.45\textwidth}
		\centering
		\includegraphics[width=\linewidth, trim=10 10 0 0, clip, page=2]{close.pdf}
		\subcaption{Recon with a closed PMT}
		\label{fig: mask}
	\end{minipage}

	\caption{Vertex reconstruction results at $r=$~\SI{260}{mm} subjected to the focus effect. The color is the count of the reconstructed vertices. (\subref{fig: all}) is the $\theta_\mathrm{v}$-$\phi_\mathrm{v}$ map of the vertex reconstruction results in spherical coordinates. The circles indicate the PMT-center directions. (\subref{fig: mask}) is the reconstruction masking out the triangle PMT, noted as ``masked''.}
\end{figure}

The opposite PMT plays an important role in the focus region. Vertex reconstruction at $r \SI{\sim 260}{mm}$ of simulated events from the Jinping prototype is shown in figure~\ref{fig: all}. The hotspots in the zenith $\theta_\mathrm{v}$ and azimuth $\phi_\mathrm{v}$ map coincide with the PMT-center directions.  Our speculation is verified by masking out one PMT in eq. \eqref{eq: MC_Likelihood}, resulting in figure~\ref{fig: mask}. The hotspot disappears around the masked PMT.

For $r>0.8r_\mathrm{LS}$, photons can reflect multiple times. Figure~\ref{fig: 5/6} takes the slice of $\theta=5\pi/6$ as an example. The superposition of multiple reflections makes the predicted PE peak at around $r>0.8r_\mathrm{LS}$, causing degeneracy.  The two local solutions of figure~\ref{fig: Likelihood2} coincides with the degenerate region of the $r_\mathrm{PMT}=\SI{900}{mm}$ in figure~\ref{fig: 5/6}.  Evident in figure~\ref{fig: Probe_PE}, such a multiple-reflection region forms a belt, unfortunately it is no longer possible to select out 3 PMTs to derive a simple criterion.

Generally speaking, degeneracy is inevitable for inhomogeneous materials. Each PMT contributes a specific degenerate region to the likelihood function.  For example, vertices around $r=0.5 r_\mathrm{LS}$ could degenerate with both the focus region and the TR region simultaneously. The situation is further complicated by the fluctuations on every PMT.  Luckily, if the degeneracy scale is less than the vertex resolution, the reconstruction result is effectively not ambiguous anymore. Higher PMT coverage gives a larger slope in the predicted PEs, narrowing the degenerate region, as illustrated by the $r_\mathrm{PMT}=\SI{850}{mm}$ line in figure~\ref{fig: 5/6}.  Therefore we recommend a small $r_\mathrm{PMT}$ at the lower limit in figure~\ref{fig: PMT_coverage}.

\subsection{Breaking of spherical symmetry}
\label{sec: symmetry}
In figure~\ref{fig: energy_0.83_z}, energy reconstruction using true vertices ``Fit(TrueV)'' is still biased when $r=$~\SI{620}{mm} at $z$~axis. Due to the outlet in figure~\ref{fig: Jinping}, only a small fraction of the photons can be detected in the outlet direction compared to the general direction.

\begin{figure}[!htbp]
	\centering
	\begin{minipage}[htb]{0.45\textwidth}
		\centering
		\includegraphics[width=\linewidth]{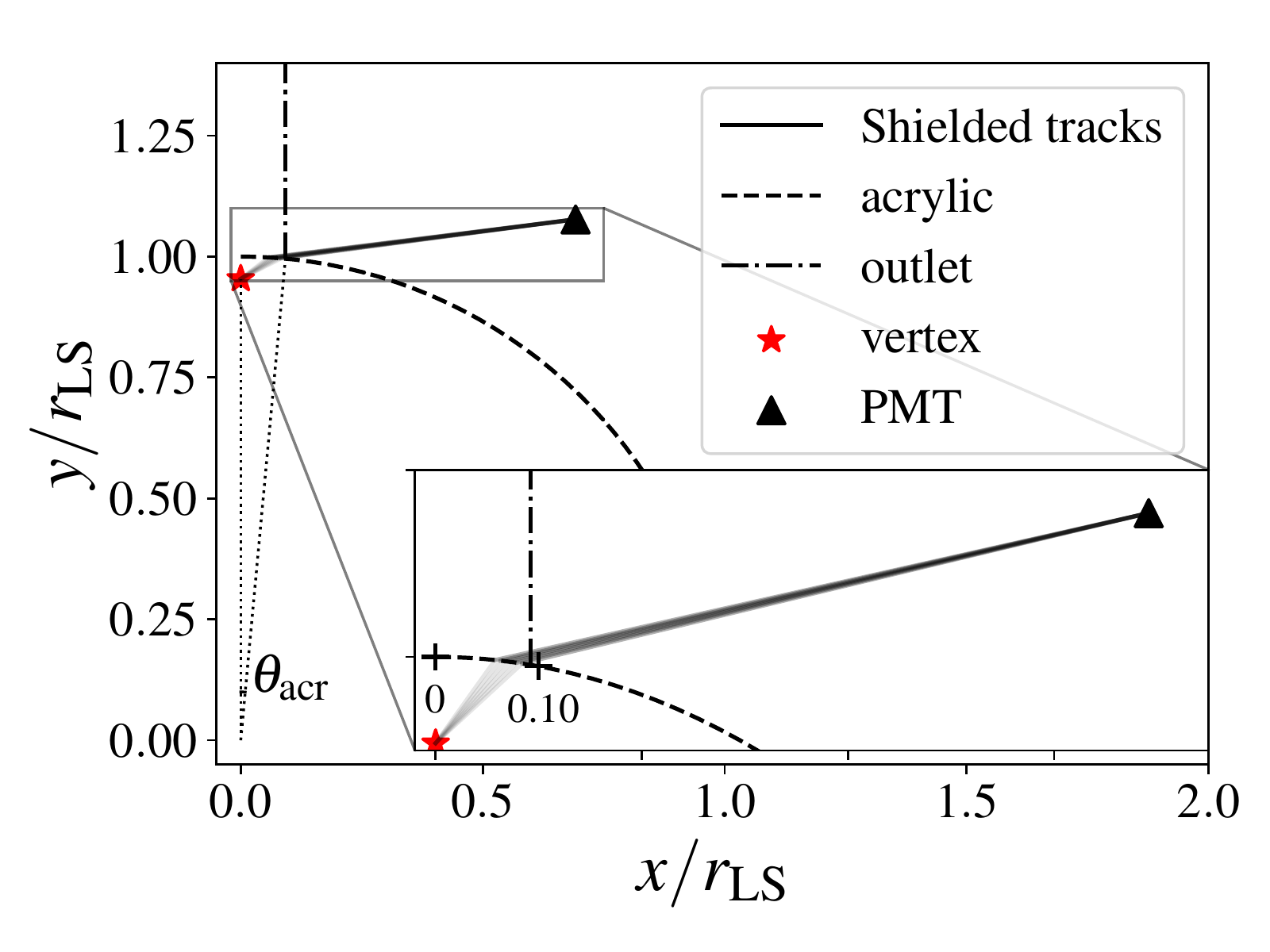}
		\subcaption{A schematic of the outlet shielding}
		\label{fig: outlet}
	\end{minipage}
	\quad
	\begin{minipage}[htb]{0.45\textwidth}
		\centering
		\includegraphics[width=\linewidth, page=1]{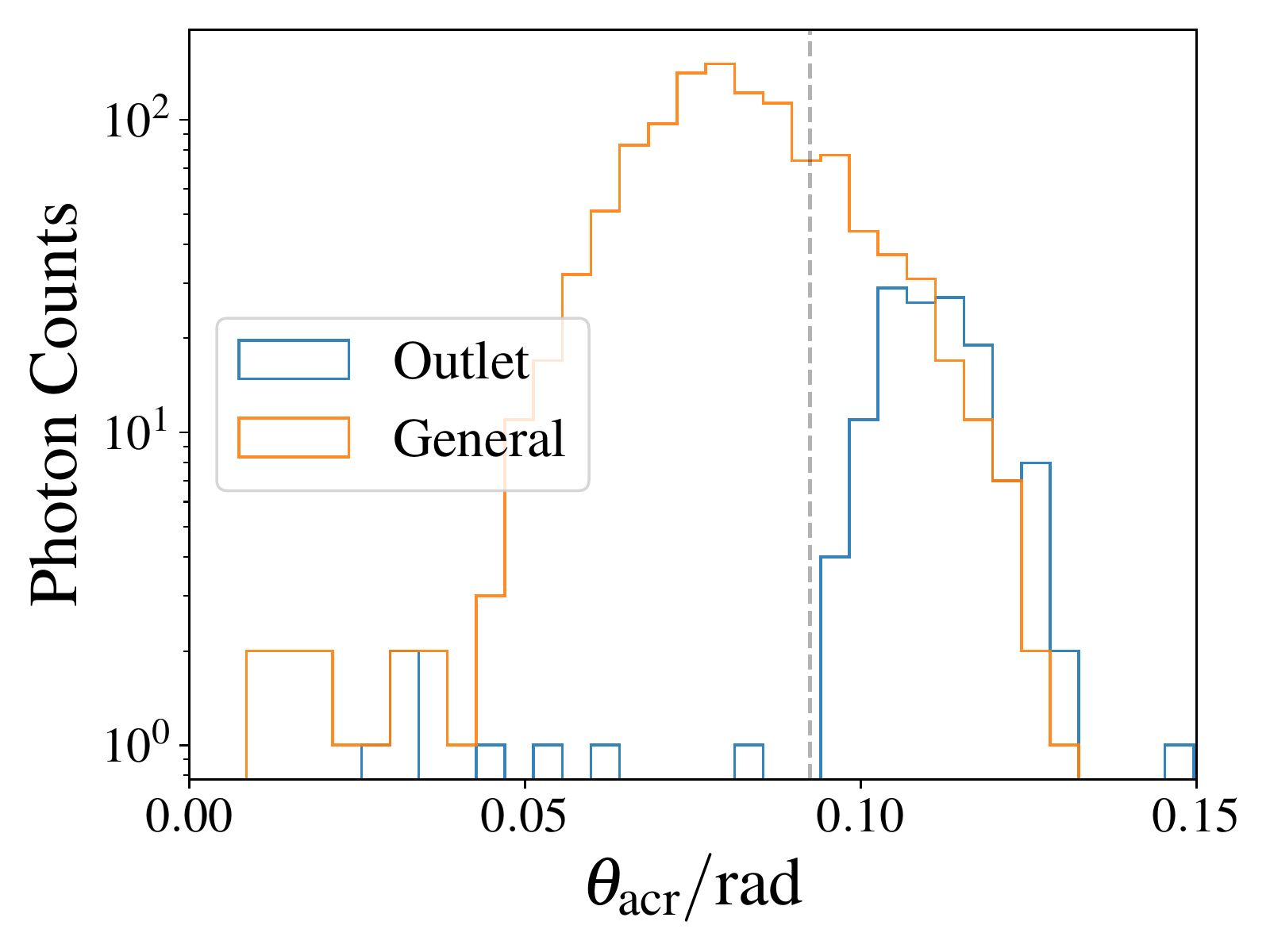}
		\subcaption{Photon tracks of general and outlet direction}
		\label{fig: track}
	\end{minipage}
	\caption{(\subref{fig: outlet}) Vertices are at $r=$ \SI{620}{mm}. The \emph{shielded tracks} are the detected tracks in the general direction but not in the outlet direction. The numbers in the zoom axis show the $\theta_\mathrm{acr}$'s value in radians. (\subref{fig: track}) The outlet region is on the left side of the grey dashed line. The distribution of the $\theta_\mathrm{acr}$ of the outlet and general directions are noted as ``Outlet'' and ``General''.}
\end{figure}

In figure~\ref{fig: outlet}, the outlet breaks spherical symmetry. Figure~\ref{fig: track} compares the $\theta_{\mathrm{acr}}$ distributions of the outlet and general directions. Photons with $\theta_\mathrm{acr} < 0.10$~rad are trapped into the outlet and seldomly detected by any PMT.

\subsection{Cosine distance for timing}
In section~\ref{subsec: cosine distance}, we ignore the timing information since PE is dominant in the Jinping Prototype. We can expand the vector $\Lambda$ to include timing for large detectors. The input is the average function $\lambda_i R_i(t)$ of the inhomogeneous Poisson process on each PMT. For continuous form, the $d_\mathrm{cos}$ is
\begin{equation}
	\begin{aligned}
		  & d_\mathrm{cos}\left[\lambda_i(\mathbf{r}_1) R_i(t; \mathbf{r}_1) \mid \lambda_i(\mathbf{r}_2) R_i(t; \mathbf{r}_2)\right]                                                                                                                                                                        \\
		= & 1- \frac{\sum_i \int \lambda_i(\mathbf{r}_1) R_i(t; \mathbf{r}_1)\lambda_i(\mathbf{r}_2) R_i(t; \mathbf{r}_2) \mathrm{d}t}{\sqrt{\sum_i \int [\lambda_i(\mathbf{r}_1) R_i(t; \mathbf{r}_1)]^2 \mathrm{d}{t}} \sqrt{\sum_i \int [\lambda_i(\mathbf{r}_2) R_i(t; \mathbf{r}_2)]^2 \mathrm{d}{t}}}.
	\end{aligned}
\end{equation}

\subsection{Upgrading Jinping Prototype}
The criterion in section~\ref{subsec: criterion} is a guideline for future detectors. Due to the space limits, PMT coverage over \SI{40}{\%} is difficult to achieve at a small radius.  Vertices in the central and the TR region make a huge difference in energy reconstruction.  If we focus on the most severe degeneracy at the detector center and \SI{600}{mm} on the $z$-axis, the main problem is that the gap between PMT1 and PMT3 is too large, as shown in figure~\ref{fig: Equiv_outlet}.  Vertices at the bisector of the 1st and 2nd closest PMTs are in the TR regions of both PMTs.  Therefore the criterion could be updated to ``the TR regions of the neighboring PMTs do not overlap''.  Compared to the criterion in section~\ref{subsec:  Cosine distance for 3-PMT case} and figure~\ref{fig: PMT_coverage} where the vertices are in the PMT-center direction, the updated minimum number of the PMTs relaxes to 1/2 in a two-dimensional slice and 1/4 in three dimensions. It shows that a minimum of 30 PMTs could meet the demand, but should be uniformly distributed.  In the current 30-PMT arrangement, 0-to-\SI{600}{mm} large degeneracy is caused by the absence of PMTs in the outlet direction, a big non-uniformity.  The Jinping Neutrino Experiment collaboration~(JNE) plans to upgrade the prototype to 60 PMTs and narrow the gaps between the PMTs around the outlet.

\section{Conclusion}
We obtain an accurate detector response model using regression, in which the number of parameters is determined by validation.  Compared to purely optics-motivated models inevitably biased at the TR regions and data-driven ones overwhelmed by the degrees of freedom, our model achieves an optimal balance.  It is shown to be unbiased in the TR regions while keeping the model complexity under control.  For \SI{2}{MeV} $e^{-}$ Jinping prototype simulation, the vertex resolution is \SI{60}{mm} at the center and \SI{20}{mm} at the TR region except those around the outlet.  The energy resolution is \SI{11}{\%}$/\sqrt{E\mathrm{[MeV]}}$. The method is confirmed to work on Jinping prototype raw data by \ce{^{214}Bi}-\ce{^{214}Po} analysis.  We believe our construction suits all the spherical detectors, especially handles the optical complexity of those with the TR.

We investigate the reconstruction degeneracy at the Jinping prototype, and confirm its origin to be the multimodality in likelihood functions.  With a set of carefully chosen approximations, we derive a necessary condition for a detector to be free from reconstruction degeneracy: the expected PE ratio between the 2 closest PMTs of any event should be less than 10.  The criterion is justified by comparing the Jinping prototype and an ideal detector setup.  That simple degeneracy criterion marks the first thorough and systematic study on the mis-reconstruction of LS detectors known to us.  It guides the upgrade of the Jinping prototype and hopefully will serve as a valuable reference for the design of PMT configurations at future LS detectors.

\section*{Acknowledgments}
We appreciate the development of the simulation tool of JSAP by Linyan Wan, Ziyi Guo and Lei Guo. Many thanks to the efforts in the waveform reconstruction by Aiqiang Zhang and Dacheng Xu. We are also grateful to Wentai Luo and Xuewei Liu for discussions on reconstruction. The PMT arrangement is inspired by Bohan Qi.  The Corresponding author would like to thank the XMASS collaboration for facilitating the idea of detector response modeling by spherical harmonics. The early seed of this paper roots in exciting discussions with Professors Shiro Ikeda, John Gregory Learned, Kai Uwe Martins and Yoichiro Suzuki.  We also thank the Jinping Neutrino Experiment collaboration for sharing the data from the Jinping 1-ton prototype. This work was supported in part by the National Natural Science Foundation of China (No. 12127808 and 12141503) and the Key Laboratory of Particle and Radiation Imaging (Tsinghua University).

\bibliographystyle{JHEP}
\bibliography{ref}

\providecommand{\href}[2]{#2}\begingroup\raggedright\begin{thebibliography}{10}

\bibitem{2002science}
{\scshape Borexino} collaboration, \emph{{Science and technology of Borexino: a
  real-time detector for low energy solar neutrinos}},
  \href{https://doi.org/10.1016/S0927-6505(01)00110-4}{\emph{Astroparticle
  Physics} {\bfseries 16} (2002) 205}.

\bibitem{PhysRevLett.100.221803}
{\scshape KamLAND} collaboration, \emph{{Precision Measurement of Neutrino
  Oscillation Parameters with KamLAND}},
  \href{https://doi.org/10.1103/PhysRevLett.100.221803}{\emph{Phys. Rev. Lett.}
  {\bfseries 100} (2008) 221803}.

\bibitem{Andringa_2016}
{\scshape SNO+} collaboration, \emph{{Current Status and Future Prospects of
  the SNO+ Experiment}}, \href{https://doi.org/10.1155/2016/6194250}{\emph{Adv.
  High Energy Phys.} {\bfseries 2016} (2016) 6194250}
  [\href{https://arxiv.org/abs/1508.05759}{{\ttfamily 1508.05759}}].

\bibitem{JUNO:2021vlw}
{\scshape JUNO} collaboration, \emph{{JUNO physics and detector}},
  \href{https://doi.org/10.1016/j.ppnp.2021.103927}{\emph{Prog. Part. Nucl.
  Phys.} {\bfseries 123} (2022) 103927}
  [\href{https://arxiv.org/abs/2104.02565}{{\ttfamily 2104.02565}}].

\bibitem{an2016neutrino}
{\scshape JUNO} collaboration, \emph{{Neutrino Physics with JUNO}},
  \href{https://doi.org/10.1088/0954-3899/43/3/030401}{\emph{Journal of Physics
  G: Nuclear and Particle Physics} {\bfseries 43} (2016) 030401}.

\bibitem{noauthor_juno_2022}
{\scshape JUNO} collaboration, \emph{{JUNO} physics and detector},
  \href{https://doi.org/10.1016/j.ppnp.2021.103927}{\emph{Progress in Particle
  and Nuclear Physics} {\bfseries 123} (2022) 103927}.

\bibitem{PhysRevLett.117.082503}
{\scshape KamLAND-Zen} collaboration, \emph{Search for majorana neutrinos near
  the inverted mass hierarchy region with kamland-zen},
  \href{https://doi.org/10.1103/PhysRevLett.117.082503}{\emph{Phys. Rev. Lett.}
  {\bfseries 117} (2016) 082503}.

\bibitem{KamLAND-Zen:2022tow}
{\scshape KamLAND-Zen} collaboration, \emph{{First Search for the Majorana
  Nature of Neutrinos in the Inverted Mass Ordering Region with KamLAND-Zen}},
  \href{https://arxiv.org/abs/arXiv:2203.02139}{{\ttfamily arXiv:2203.02139}}.

\bibitem{SNO:2021xpa}
{\scshape SNO+} collaboration, \emph{{The SNO+ experiment}},
  \href{https://doi.org/10.1088/1748-0221/16/08/P08059}{\emph{JINST} {\bfseries
  16} (2021) P08059} [\href{https://arxiv.org/abs/2104.11687}{{\ttfamily
  2104.11687}}].

\bibitem{beacom_letter_2017}
J.~F. Beacom, S.~Chen, J.~Cheng, S.~N. Doustimotlagh, Y.~Gao, S.-F. Ge et~al.,
  \emph{Letter of {Intent}: {Jinping} {Neutrino} {Experiment}},
  \href{https://doi.org/10.1088/1674-1137/41/2/023002}{\emph{Chinese Phys. C}
  {\bfseries 41} (2017) 023002}.

\bibitem{Liu_2018}
Q.~Liu, M.~He, X.~Ding, W.~Li and H.~Peng, \emph{{A vertex reconstruction
  algorithm in the central detector of JUNO}},
  \href{https://doi.org/10.1088/1748-0221/13/09/t09005}{\emph{Journal of
  Instrumentation} {\bfseries 13} (2018) T09005–T09005}.

\bibitem{Kim:2012cuq}
H.~S. Kim, \emph{{Finding an Event Vertex by Using a Weighting Method at
  RENO}}, \href{https://doi.org/10.3938/NPSM.62.631}{\emph{New Phys. Sae Mulli}
  {\bfseries 62} (2012) 631}.

\bibitem{li2021event}
Z.~Li, Y.~Zhang, G.~Cao, Z.~Deng, G.~Huang, W.~Li et~al., \emph{Event vertex
  and time reconstruction in large-volume liquid scintillator detectors},
  \href{https://doi.org/10.1007/s41365-021-00885-z}{\emph{Nuclear Science and
  Techniques} {\bfseries 32} (2021) 1}.

\bibitem{CHOOZ:2002qts}
{\scshape CHOOZ} collaboration, \emph{Search for neutrino oscillations on a
  long baseline at the {CHOOZ} nuclear power station},
  \href{https://doi.org/10.1140/epjc/s2002-01127-9}{\emph{Eur. Phys. J. C}
  {\bfseries 27} (2003) 331}
  [\href{https://arxiv.org/abs/hep-ex/0301017}{{\ttfamily hep-ex/0301017}}].

\bibitem{Back_2012}
{\scshape Borexino} collaboration, \emph{{Borexino calibrations: hardware,
  methods, and results}},
  \href{https://doi.org/10.1088/1748-0221/7/10/p10018}{\emph{Journal of
  Instrumentation} {\bfseries 7} (2012) P10018–P10018}.

\bibitem{Tajima:2003zz}
O.~Tajima, \emph{Measurement of electron anti-neutrino oscillation parameters
  with a large volume liquid scintillator detector, {KamLAND}},  2003.

\bibitem{GALBIATI2006700}
C.~Galbiati and K.~McCarty, \emph{Time and space reconstruction in optical,
  non-imaging, scintillator-based particle detectors},
  \href{https://doi.org/10.1016/j.nima.2006.07.058}{\emph{Nuclear Instruments
  and Methods in Physics Research Section A: Accelerators, Spectrometers,
  Detectors and Associated Equipment} {\bfseries 568} (2006) 700}.

\bibitem{renocollaboration2010reno}
{\scshape RENO} collaboration, \emph{{RENO: An Experiment for Neutrino
  Oscillation Parameter $\theta_{13}$ Using Reactor Neutrinos at Yonggwang}},
  \href{https://arxiv.org/abs/arXiv:1003.1391}{{\ttfamily arXiv:1003.1391}}.

\bibitem{Huang:2021baf}
G.~Huang, Y.~Wang, W.~Luo, L.~Wen, Z.~Yu, W.~Li et~al., \emph{{Improving the
  energy uniformity for large liquid scintillator detectors}},
  \href{https://doi.org/10.1016/j.nima.2021.165287}{\emph{Nucl. Instrum. Meth.
  A} {\bfseries 1001} (2021) 165287}
  [\href{https://arxiv.org/abs/2102.03736}{{\ttfamily 2102.03736}}].

\bibitem{qian2021vertex}
Z.~Qian, V.~Belavin, V.~Bokov, R.~Brugnera, A.~Compagnucci, A.~Gavrikov et~al.,
  \emph{Vertex and energy reconstruction in {JUNO} with machine learning
  methods}, \href{https://doi.org/10.1016/j.nima.2021.165527}{\emph{Nuclear
  Instruments and Methods in Physics Research Section A: Accelerators,
  Spectrometers, Detectors and Associated Equipment} (2021) 165527}.

\bibitem{hastie01statisticallearning}
T.~Hastie, R.~Tibshirani and J.~Friedman, \emph{The Elements of Statistical
  Learning: Data Mining, Inference, and Prediction}. Springer New York, 2013.

\bibitem{10.2307/2344614}
J.~A. Nelder and R.~W.~M. Wedderburn, \emph{Generalized linear models},
  {\emph{Journal of the Royal Statistical Society. Series A (General)}
  {\bfseries 135} (1972) 370}.

\bibitem{davino2013quantile}
C.~Davino, M.~Furno and D.~Vistocco, \emph{Quantile Regression: Theory and
  Applications}. Wiley, 2013.

\bibitem{noll1976zernike}
R.~J. Noll, \emph{Zernike polynomials and atmospheric turbulence},
  \href{https://doi.org/10.1364/JOSA.66.000207}{\emph{JOsA} {\bfseries 66}
  (1976) 207}.

\bibitem{agostinelli2003geant4}
S.~Agostinelli, J.~Allison, K.~Amako, J.~Apostolakis, H.~Araujo, P.~Arce
  et~al., \emph{Geant4—a simulation toolkit},
  \href{https://doi.org/https://doi.org/10.1016/S0168-9002(03)01368-8}{\emph{Nuclear
  Instruments and Methods in Physics Research Section A: Accelerators,
  Spectrometers, Detectors and Associated Equipment} {\bfseries 506} (2003)
  250}.

\bibitem{WANG201781}
Z.~Wang, Y.~Wang, Z.~Wang, S.~Chen, X.~Du, T.~Zhang et~al., \emph{{Design and
  analysis of a 1-ton prototype of the Jinping Neutrino Experiment}},
  \href{https://doi.org/10.1016/j.nima.2017.03.007}{\emph{Nuclear Instruments
  and Methods in Physics Research Section A: Accelerators, Spectrometers,
  Detectors and Associated Equipment} {\bfseries 855} (2017) 81}.

\bibitem{gonzalez2010measurement}
{\'A}.~Gonz{\'a}lez, \emph{Measurement of areas on a sphere using {Fibonacci}
  and latitude--longitude lattices},
  \href{https://doi.org/10.1007/s11004-009-9257-x}{\emph{Mathematical
  Geosciences} {\bfseries 42} (2010) 49}.

\bibitem{guo2019slow}
Z.~Guo, M.~Yeh, R.~Zhang, D.-W. Cao, M.~Qi, Z.~Wang et~al., \emph{Slow liquid
  scintillator candidates for {MeV}-scale neutrino experiments},
  \href{https://doi.org/10.1016/j.astropartphys.2019.02.001}{\emph{Astroparticle
  Physics} {\bfseries 109} (2019) 33}.

\bibitem{zhao2022measurement}
{\scshape JNE} collaboration, \emph{Measurement of muon-induced neutron
  production at {China Jinping Underground Laboratory}},
  \href{https://doi.org/10.1088/1674-1137/ac66cc}{\emph{Chinese Physics C}
  (2022) }.

\bibitem{kraft1988software}
D.~Kraft, \emph{A software package for sequential quadratic programming}. Tech
  Rep DFVLR-FB 88-28, 1988.

\bibitem{abusleme2020tao}
{\scshape JUNO} collaboration, \emph{{TAO Conceptual Design Report: A Precision
  Measurement of the Reactor Antineutrino Spectrum with Sub-percent Energy
  Resolution}},  \href{https://arxiv.org/abs/arXiv:2005.08745}{{\ttfamily
  arXiv:2005.08745}}.

\bibitem{abe2013xmass}
{\scshape XMASS} collaboration, \emph{{XMASS detector}},
  \href{https://doi.org/10.1016/j.nima.2013.03.059}{\emph{Nuclear Instruments
  and Methods in Physics Research Section A: Accelerators, Spectrometers,
  Detectors and Associated Equipment} {\bfseries 716} (2013) 78}.

\bibitem{abe_development_2019}
{\scshape XMASS} collaboration, \emph{Development of low radioactivity
  photomultiplier tubes for the {XMASS}-{I} detector}, {\emph{Nuclear
  Instruments and Methods in Physics Research Section A: Accelerators,
  Spectrometers, Detectors and Associated Equipment} {\bfseries 922} (2019)
  171}.

\bibitem{richardson1972bayesian}
W.~H. Richardson, \emph{Bayesian-based iterative method of image restoration},
  \href{https://doi.org/10.1364/JOSA.62.000055}{\emph{JoSA} {\bfseries 62}
  (1972) 55}.

\bibitem{xu2021towards}
D.~C. Xu, B.~D. Xu, E.~J. Bao, Y.~Y. Wu, A.~Q. Zhang, Y.~Y. Wang et~al.,
  \emph{Towards the ultimate {PMT} waveform analysis for neutrino and dark
  matter experiments},
  \href{https://doi.org/10.1088/1748-0221/17/06/p06040}{\emph{Journal of
  Instrumentation} {\bfseries 17} (2022) P06040}.

\bibitem{Kullback1951}
S.~Kullback and R.~A. Leibler, \emph{On information and sufficiency},
  \href{https://doi.org/10.1214/aoms/1177729694}{\emph{The annals of
  mathematical statistics} {\bfseries 22} (1951) 79}.

\bibitem{10.2307/2337385}
A.~Basu, I.~R. Harris, N.~L. Hjort and M.~C. Jones, \emph{Robust and efficient
  estimation by minimising a density power divergence},
  \href{https://doi.org/10.1093/biomet/85.3.549}{\emph{Biometrika} {\bfseries
  85} (1998) 549}.

\bibitem{alimonti2000light}
{\scshape Borexino} collaboration, \emph{Light propagation in a large volume
  liquid scintillator},
  \href{https://doi.org/10.1016/S0168-9002(99)00961-4}{\emph{Nuclear
  Instruments and Methods in Physics Research Section A: Accelerators,
  Spectrometers, Detectors and Associated Equipment} {\bfseries 440} (2000)
  360}.

\end{thebibliography}\endgroup
\end{document}